\documentclass[letterpaper,twocolumn,10pt]{article}
\usepackage{usenix,epsfig,endnotes}
\usepackage{epsfig,endnotes}

\usepackage{graphicx}
\usepackage{outlines}
\usepackage{url}
\usepackage{tikz}
\usetikzlibrary{arrows.meta,positioning,fit,calc}
\usepackage{float}
\usepackage{amsmath}
\usepackage{xcolor}
\usepackage{siunitx}
\usepackage{xspace}
\usepackage{placeins} 
\usepackage{stfloats} 
\newcommand{\liha}{\textsc{LiHA}\xspace}

\begin{document}

\title{\liha: Worse is Better for Lightweight Hash-based Source Authentication } 

\author{
    {\rm Negin Dehghanchaleshtori}\\
    Rutgers University 
    \and
    {\rm Homit Dalia}\\
    Rutgers University 
    \and
    {\rm Himanshu Chandra}\\
    Rutgers University 
    \and 
    {\rm Rohan Tripathi}\\
    Rutgers University 
    \and
    {\rm Marc E. Fiuczynski}\\
    Rutgers University
    \and 
    {\rm Richard P. Martin}\\
    Rutgers University
    \and 
    {\rm Rutgers University Department of Computer Science Technical Report DCS-TR-764}\\
    {\rm October, 2026} \\
}

\date{October, 2026}

\maketitle

\begin{abstract}
The omission of source IP authentication remains a fundamental vulnerability in the Internet. Existing approaches rely on routing plausibility, topology assumptions, per-flow state, or wide deployment. These requirements have kept strong source authentication out of the data plane.

As simplicity is the key to deployability, we present \liha, a lightweight source authentication mechanism built around operations that routers already perform efficiently. \liha binds secret keys to source prefixes using longest prefix match and authenticates each packet with a single message authentication code (MAC). The data plane requires no handshakes, per-flow state, or per-packet control plane access. A sender performs one prefix lookup and one fixed-size hash. The verifier repeats the same operations and forwards the packet only if the MAC matches.

\liha is designed for incremental deployment. It primarily protects against off-path attackers that spoof source prefixes they do not control. An on-path attacker can replay valid authenticated packets until the key rotates. \liha does not eliminate every denial-of-service threat. Instead, it confines accepted traffic to authenticated source prefix space within participating networks. We argue that simpler source authentication with clear limits is more likely to be deployed.

We implemented \liha in three environments: as a set of user-space libraries, as eBPF programs, and in an FPGA. 
We found that the user-space implementation had very high overheads. The eBPF implementation got reasonable single thread performance even
with a software based cryptographic hash. We found that only the FPGA could obtain 100 Gbps single thread performance, but that was only when
bypassing the kernel using a DPDK user-level networking stack. 

\end{abstract}

\section{Introduction}
\label{sec:Introduction}

The foundational Internet architecture was designed without source
authentication. In the beginning this was reasonable. Networks were small and trusted,
router resources were limited, and forwarding was optimized for efficiency. Today's Internet is radically different. It is ubiquitous and adversarial; yet
the network layer still provides no built-in way to verify
a transmitter's identity thus opening up the network to
a host of problems including Distributed Denial of Service (DDoS)
attacks, reflection and amplification attacks, Internet-wide scanning
campaigns, and session hijacking.

The community has addressed this problem; however, many solutions subscribe to the perfect authentication fallacy. Topological approaches include BCP 38~\cite{bcp38} and unicast Reverse Path Forwarding (uRPF)~\cite{bcp84}. Stronger session security is the goal of the Internet Protocol Security (IPsec) protocol. Path authentication approaches include Passport~\cite{liu2008passport} and the Accountable Internet Protocol (AIP)~\cite{andersen2008accountable}.

The failure of adopting many of these approaches stems from the problem of managing changes in evolutionary systems. Such systems grow with changes constrained to maintaining the functionality of existing, working interfaces. Realizing changes in evolutionary systems can thus follow one of three paths: (1) preserve the existing structure and build on top, as is seen with the deployment of TLS over TCP, (2) radical interface changes can start small and grow over time, sometimes requiring decades and never fully replacing the old interfaces, such as with IPv6, or (3) slowly add features while maintaining compatibility, as exemplified by Ethernet. A defense that requires a new Internet is no defense at all. 

Prior approaches to authentication mostly ignored evolutionary constraints. They added headers, changed the protocol, added too much state or message exchanges, or added administrative overhead.  Any deployable change must balance minimizing costs while maximizing efficacy. Much security work focuses on maximizing efficacy and ignores deployment costs. Much networking research maximizes line-rate performance and ignores other factors. Proposed changes subscribing to such maximalist positions are doomed to failure for any evolutionary-constrained system, the Internet included.  

Another way to characterize the evolution constraint is as an inverse of Metcalfe's Law. Recall that Metcalfe's law states that network value scales as
$N^2$: the value of the network increases as the square of the number of communicating parties~\cite{metcalfe2013metcalfe}. However, a corollary law is that deployment taxes also scale with $N^2$; changes become quadratically more difficult as the network scales. 

We take a radically different approach from prior art and embrace the constraint that any
addition must be compatible and make minimal demands on the existing framework. 
Because minimal change drives adoption, a partial compatible solution is better than a complex, incompatible,
yet complete one. Such a value system was articulated as the ``New Jersey Approach'' in Gabriel's essay "The Rise of Worse is Better"
on system design~\cite{gabriel1991worse}.

Our minimalist strategy is called Lightweight Hash-based Source Authentication (\liha{}). Our goal is to 
add an authentication primitive in a manner that makes the least demands on the existing Internet while still 
adding meaningful packet authentication. Following the New Jersey approach it makes these design choices: 
\vspace{-0.02in}
\begin{itemize}
\item Move key distribution out of the communication path to external servers.
\item Rely on rapid key rotation to mitigate replay attacks.
\item Limit the concurrent state to only the secret key. Do not use nonces or other state.
\item Use existing IP options rather than add new headers or fields.
\item Use a strong single hash function rather than the double-hashing used in standard HMAC.
\item Hash only over selected IP header fields and key, using the source IP prefix as the security context. 
\end{itemize} 

Unlike every other protocol based approach \liha{} requires an external server for the
sender and verifier to communicate the secret keys. This greatly simplifies the protocol,
the system, and allows for much less state in the networking equipment. We show external
key exchange and updates can be performed in under 1 second in practice.

We show that \liha{} is robust to packet alteration, lightweight, performant,
and straightforward to implement.  We prove its minimalism and simplicity by creating multiple
implementations at different points in the network: on hosts at the user-level, in the kernel, and inside
a network interface card. 
This implementation flexibility restores decentralized operator autonomy. It allows both large and small ISPs to 
offer filtering services that remove malicious traffic rather than disperse it behind expensive and complex
anycast networks,
thus reclaiming resources that would otherwise be wasted on attack traffic.
Its simplicity allows any organization that manages IP networks to implement filtering at low cost. 
The minimalist approach preserves full compatibility with
existing infrastructure while satisfying the three critical
constraints of high-speed forwarding: near-stateless operation,
portability across environments, and incremental deployability without universal adoption.

We make the following contributions:
\begin{itemize}
\item \textbf{Diverse Prototyping.} We implement \liha{} across three prototypes—a pure user-level application, an eBPF module, and an FPGA implementation— demonstrating that \liha{} requires minimal modifications to operating systems, routers, and existing networking equipment.
\item \textbf{High-Speed Performance.} We achieve 100 Gbps line-rate source authentication in our FPGA implementation and deliver performance comparable to single-thread execution using eBPF, including the forwarding path source lookup.
\item \textbf{Qualitative Evaluation.} We evaluate the development effort and architectural complexity of integrating \liha{} at different layers of the network stack.
\item \textbf{Practical Mitigation.} We demonstrate \liha{}'s effectiveness by mitigating an emulated DDoS attack, proving that it preserves internal network bandwidth even when the inbound local area network (LAN) link approaches saturation.
\end{itemize}

The remainder of this paper is organized as follows. Section~\ref{sec:related} reviews
related work. Section ~\ref{sec:threat} defines our threat model. Section~\ref{sec:design} presents the \liha design.  Section~\ref{sec:implementation} describes our implementations and presents our experimental results. Section ~\ref{sec:limitations} discusses limitations of \liha design, and Section ~\ref{sec:conclusion} concludes the paper.

\section{Related Work}
\label{sec:related}
Existing authentication mechanisms can be grouped into five categories: filter-based, behavioral checks, traceback after attack, access validation, and cryptographic authentication. We present representative works in each category as space limitations prevent a full accounting of each approach. 

Filtering based approaches ask if an address logically belongs where it is observed. Ingress filtering ~\cite{bcp38} drops packets whose source address should not appear on a given interface. Unicast Reverse Path Forwarding (uRPF) ~\cite{bcp84} extends this concept by confirming the source can be reached via the arrival interface. However, its strict mode breaks under asymmetric routing, while its loose mode trades precision for deployability.
Route-based filters ~\cite{rbf} and inter-domain packet filters ~\cite{idpf} make similar decisions using routing information, often derived from BGP. These approaches check whether a source is plausible, while \liha checks whether the packet carries a valid MAC for that source prefix. These techniques are useful and widely understood, but they become less effective when deployment is incomplete, routing information is out of date, or packets take different paths in each direction.

Another group of approaches looks for signs of spoofing in packet behavior. Hop-count filtering, for example, compares a packet’s TTL with the expected distance from its claimed source~\cite{hcf}. In general, these methods check whether a packet followed a plausible path. However, a careful attacker can often adjust the packet to avoid detection.

Traceback approaches focus on identifying the source of an attack after it has occurred. They do this by adding path information to packets ~\cite{traceback}, keeping state at routers ~\cite{snoeren}, or sending separate control messages ~\cite{icmptraceback}. These methods can help identify the source of an attack. However, they do not block spoofed packets when they arrive. Their effectiveness also depends on how many routers along the path support them.

Closer to the source, access-validation methods bind an address to the point it was assigned. Access validation checks whether a host is authorized to use the address it claims. SAVI~\cite{savi} links each IP address to a host’s network connection point. This prevents hosts on the same link from using addresses assigned to others. SAVA~\cite{sava} extends this idea to larger networks by applying checks at the access, intra-AS, and inter-AS levels. These approaches provide stronger validation than prefix-level filtering. However, they need extra state and coordination to keep source address information up to date. \liha shares the same goal of stopping spoofed traffic close to its source. Unlike these approaches it does not maintain any state about where a host is connected. Instead, it makes its decision using a small set of header fields and a single prefix-to-key lookup. This simple design allows \liha to run in routers, NICs, or middleboxes. The IETF SAVNET effort addresses the same problem using routing information, and its mechanisms can be used alongside the cryptographic checks performed by \liha.

Cryptographic authentication provides the strongest protection against source spoofing. \liha{} belongs to this category. Many cryptographic systems protect a connection rather than each packet. IPsec~\cite{ipsec}, WireGuard~\cite{wireguard}, OpenVPN~\cite{openvpn}, TLS 1.3~\cite{tls13}, QUIC~\cite{quic}, and MACsec~\cite{macsec} all authenticate traffic. Generally they need per-flow or per-peer state, a handshake, and replay protection.
Passport~\cite{liu2008passport} is closer to \liha because it authenticates individual packets. They allow receivers, and sometimes routers along the path, to verify a packet’s claimed source. Passport also showed that this can be done without maintaining per-flow state. The main difference between Passport and \liha{} is their deployment model. Passport distributes per-AS keys through BGP and places one authenticator for each AS on the path into the packet. This changes the packet format and increases the amount of authentication data carried by each packet. \liha{} takes a simpler approach. It uses one key per source prefix and distributes keys out of band over standard TLS. Each packet carries a single MAC that is checked when the packet arrives. The result is an immediate accept or drop decision.
The two systems also operate at different levels. Passport authenticates traffic at AS granularity. \liha{} associates keys with source prefixes, which operators already use for routing and policy. This provides finer isolation for customers or tenants within the same AS.

SCION and LightningFilter ~\cite{scion, lightningfilter} provide another useful comparison. Their implementations show that cryptographic checks can be performed at high packet rates, with LightningFilter demonstrating filtering at 100~Gbps. \liha{} does not claim that fast authentication is new. Its contribution is a design that fits within the existing Internet architecture. SCION requires a new path-aware network architecture, while LightningFilter relies on DRKey for key establishment. \liha{}, by contrast, works with standard IPv6 and SRv6 packets. It uses TLS to distribute keys.

SRv6 already provides a field for packet authentication. RFC~8754~\cite{srh} defines an HMAC TLV in the Segment Routing Header. However, the standard does not provide a scalable way to distribute and manage the required keys. \liha{} provides this missing key-management layer. It therefore uses an existing packet format rather than introducing a new one.

\section{Threat Model}
\label{sec:threat}

\liha takes a different design philosophy than most works: it is not the network's job to stop all attacks, rather it should make every packet attributable, filterable, billable and prosecutable. When prefixes become easier to trace, operators can revoke the corresponding keys or apply targeted filtering. If attackers use a VPN or proxy, the traffic is associated with that service's egress prefix.

The threat we are concerned with is strictly authentication: packets generated by a machine on the Internet
which has previously authenticated its public key with a key server, using an out-of-band mechanism,
should be allowed. Other traffic should be redirected, tagged or dropped. Below are the classes of threats
we address: 

{\bf Compromised Host.}
A compromised host can generate packets with valid signatures using arbitrary source addresses, but not
arbitrary keys. That is, it still must use a secret key given to it by the keyserver.
We refer to these hosts as host attackers. 

{\bf Compromised Router.}
A compromised router or middlebox can eavesdrop trafﬁc sent by legitimate sources
at its network location and replay that trafﬁc from other, compromised host locations.
We refer to these attacks as monitor attacks. 

{\bf Compromised Keyserver.}
A compromised key server can create secret keys under the control of the attacker. Such an attacker
could give different secret keys legitimate transmitters and receivers trying to communicate,
as well as duplicate or invalid keys. 

{\bf Cooperative Attackers.}
Cooperative attackers can control both a compromised keyserver as well as a compromised host or
compromised router. Such a cooperative attack can generate false secret keys as well as give those
to compromised hosts and routers.

Assuming single levels of compromise in the hierarchy, \liha provides complete protection from off-path spoofing. An on-path observer can mint attack traffic until the next key rotation, and then that prefix can be blocked by the key server. A compromised host can generate attack traffic. However, depending on the size of the verifier's prefix table and communication with the key server, the matching prefix could be reduced to just a single compromised host. A compromised verifier presents a larger version of the compromised host problem; the entire range of the verifier's prefix would have to dropped. A compromised key server results in a total failure. The goal is not to eliminate collateral damage entirely, but to keep it scaled to the level of the attacker.

\section{\liha{} Design}
\label{sec:design}

\begin{figure*}[t!]
    \centering
    \includegraphics[width=0.75\textwidth]{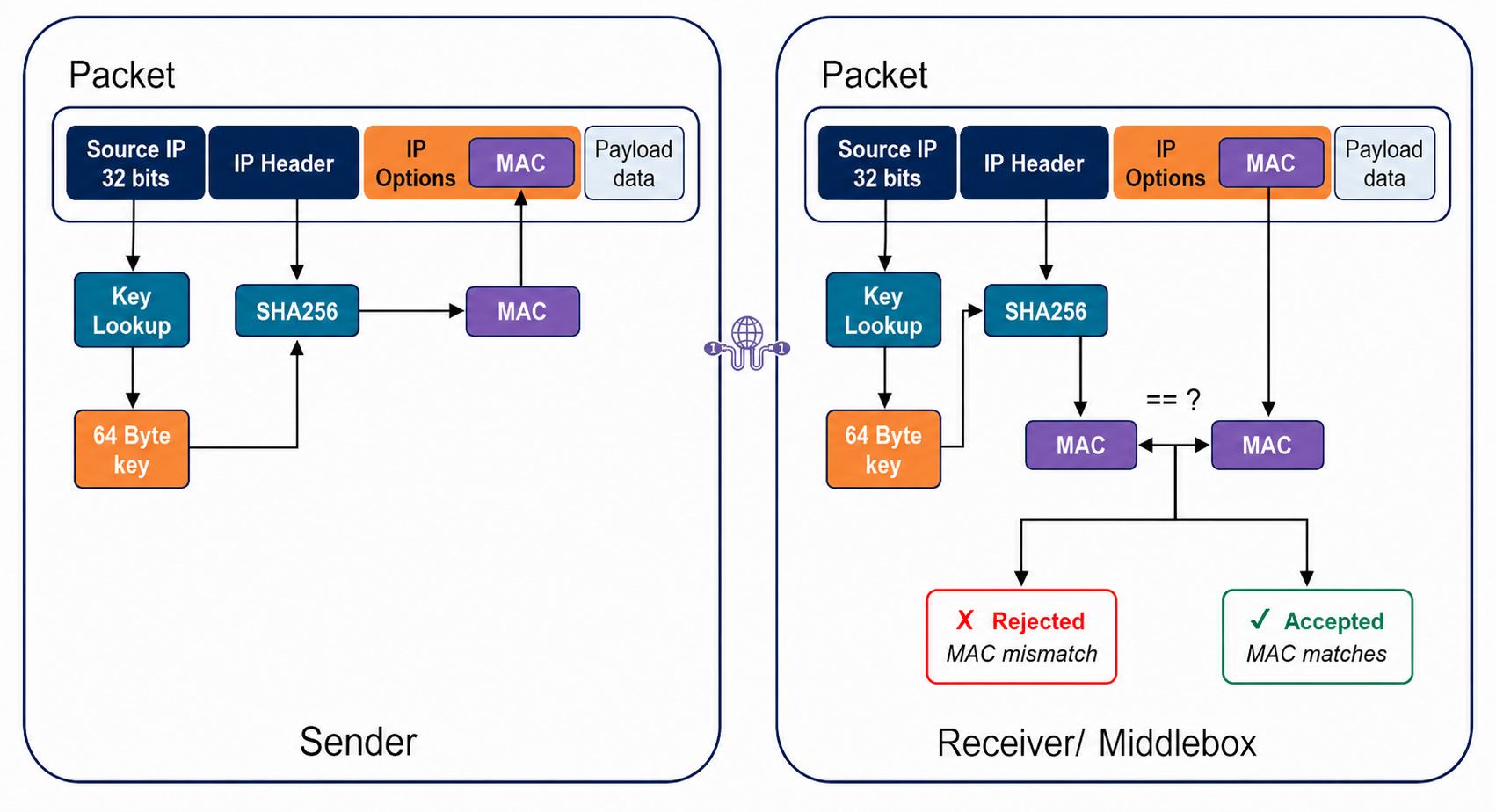}
    \caption{End-to-end \liha operation.}
    \label{fig:designmodel}
\end{figure*}

In this section, we present the \liha design. We first describe the overall approach, and then describe three implementations; an FPGA one in
Section~\ref{sec:fpga}, one in userspace in Section~\ref{sec:userspace}, and Section~\ref{sec:ebpf} describes an eBPF one. 

\liha{} uses minimal state, in-path authentication mechanism. For each outgoing packet, the sender computes a SHA-256 value over selected IP header fields with appended key, then inserts the signature into the \texttt{IP Options} field. 
A downstream authenticating node verifies the packet using the same IP header fields and key. The design requires no handshake, no connection state, and no per-flow table. 

At the sender, \liha runs a Longest Prefix Match (LPM) on the packet's source address to retrieve a 64-byte key. The key server must ensure a transmitter and receiver share a key, however, there can be different ones for each direction in the path. A transmitter computes the SHA-256 over immutable header fields with the key appended. 
The resulting message authentication code (MAC) is placed in the packet’s IP Options field.
A verifier, which may be an end host, NIC, router, or middlebox, performs the same prefix lookup and hash computation. It then compares the computed value with the MAC carried in the packet. As shown in Equation ~\ref{eq:verification}, the packet is accepted when the two values match and dropped otherwise. The design requires no handshake, connection state, or per-flow table. Figure ~\ref{fig:designmodel} shows the complete end-to-end \liha operation.

\begin{equation}
\mathrm{verify}(p):
\quad
\mathrm{MAC}_p
\stackrel{?}{=}
\mathrm{SHA256}\!\left(
\mathrm{immutable}(p) \,\|\, K_{\mathrm{LPM}(\mathrm{src})}
\right).
\label{eq:verification}
\end{equation}

where $\mathrm{immutable}(p)$ denotes the immutable IP header fields. $K_{\mathrm{LPM}(\mathrm{src})}$ is the secret key associated with the longest prefix match of the packet's source address. The remainder of this section explains the main design decisions behind \liha.

\subsection{Keys bind to prefixes}
Unlike almost all other approaches, \liha binds keys to source
prefixes rather than flows or hosts. A single key can therefore authenticate all traffic from that prefix without maintaining per-host or per-flow state. A radix trie holds a mapping between a prefix and its key. 

In our prototype, nodes fetch their keys from a trusted external service, HashiCorp Vault ~\cite{vault}, over mutually authenticated TLS. Keys are fetched at startup and during rotation, so the forwarding plane does not participate in the exchange. A complete key rotation takes approximately 30~ms. To prevent packet drops during a swap, the forwarding pipeline temporarily accepts both the old and new keys. After a period long enough to cover control plane and network delays, the old key is removed. This process does not introduce any per-packet state.

Managing the keys out of the forwarding plane simplifies deployment. The key service can be a single web server or a distributed, anycast-fronted system without changing the data path. Access control can be handled by the identity and access management layer instead of being embedded in routers. During an attack, operators can update prefix and key rules at the key service or network edge without changing the fast path.

Prefix-based keying works well when each prefix belongs to a clear trust domain. For example, all traffic from a customer’s prefix can share one key. When several tenants use the same prefix, separate prefixes are needed to authenticate them independently. This follows the same general approach operators already use when assigning prefixes for routing and access control.

This alignment is not always possible. Some IPv4 networks place many tenants behind a shared NAT and a single prefix. In such cases, assigning a separate key to every tenant would require address changes or additional mechanisms. \liha therefore authenticates the NAT boundary rather than each tenant. This limitation is similar to the attribution problems already created by shared NAT.

\subsection{Single Hash}
\liha uses a single SHA-256 computation with the secret key appended to the message. This design is more efficient than traditional HMAC, which requires two hash computations. HMAC uses the second hash partly to protect against attacks such as length extension~\cite{hmac}. \liha instead uses a secret-suffix construction, written as $H(M || K)$. The key appears after the message rather than before it. As a result, a length-extension attack cannot be applied without knowing the secret key. The attacker would need to extend the input beyond a value that is not known.

A remaining concern is collision resistance. Finding a useful SHA-256 collision would require about $2^{128}$ operations. \liha also hashes a fixed-length set of header fields, giving an attacker little control over the input. An attacker without the key may also try to guess the MAC. \liha carries the full 256 bit of SHA-256 output in the packet. A blind attacker succeeds with probability $2^{-256}$ per attempt. 

The deeper reason for using one hash is pipelining. HMAC has a serial dependency because the outer hash cannot begin until the inner hash finishes. This requires more pipeline stages or hardware on an FPGA and more CPU cycles on end hosts. A single SHA-256 pass avoids this extra work.

This type of keyed hashing has already been used in networking protocols. RIPv2, specified in RFC~2082~\cite{ripv2}, and BFD, specified in RFC~5880~\cite{bfd}, use similar methods to authenticate routing messages. \liha follows the same general approach. If SHA-256 is no longer considered secure, it can be replaced without changing the rest of the packet processing design.

\subsection{Minimal State}
\liha keeps no per-flow state in the forwarding plane. The sender and verifier store only the prefix to a key table. This keeps memory use independent of the number of active senders and allows the same design to run in software, NICs, and router hardware. 

The main tradeoff is replay protection. A captured packet remains valid until its key expires. More frequent key rotation reduces this window. A future extension could also include a coarse time epoch in the authenticated data. The verifier could accept the current and previous epochs to tolerate clock skew and packets crossing an epoch boundary. This would require no per-sender state. We leave this out of the current design to keep \liha simple. \liha therefore focuses mainly on off-path attackers that spoof source prefixes they do not control.

\subsection{MAC Placement}
The IPv4 prototype carries the MAC in the IP Options field. This avoids introducing a new header or changing the IPv4 packet format. IP Options
are not ideal for large deployment because some routers process
them on a slow path or drop them. In controlled networks, however,
operators can configure the forwarding path to accept \liha packets. Our
evaluation shows that the added option does not prevent the hardware
implementation from operating near the line rate.

IPv6 and SRv6 provide a cleaner deployment path. The Segment Routing
Header already defines an HMAC TLV that can carry authentication data
(RFC~8754~\cite{srh}). Networks that already use SRv6 can therefore add
\liha without defining another packet header. IPv4 traffic can also be
carried through an SRv6 domain, although the encapsulation adds roughly
40--64~bytes per packet. This fixed overhead reduces usable bandwidth,
but may be reasonable when the goal is to stop much larger spoofed or
amplified attacks.

\subsection{Sender Steps}
The sender is responsible for constructing the authenticated packet before transmission. The operation executes inline in the transmit path and consists of three steps.\\
\textbf{Key lookup.} The sender extracts the source IP address and performs an LPM lookup in the local radix tree. This returns the 64 Byte key for the longest prefix matching. If no prefix matches, the source address is not part of the \liha{} deployment. In that case, the packet is sent without authentication. This allows incremental deployment and does not disrupt traffic from non participating flows.\\
\textbf{Signature computation.} After obtaining the key, the sender computes the signature running
a SHA-256 hash on selected IP header fields appended with the key. It only uses header fields that stay unchanged from end to end, so sets the TTL and checksum to zero. This makes verification deterministic,
even when packets pass through middleboxes.\\
\textbf{Signature embedding.} The sender places the signature in the \texttt{IP Options} field and sends the packet normally.

\subsection{Receiver Validation}
An authenticating receiver, whether it is the end host or an intermediate middlebox, validates each packet using the same basic logic as the sender.\\
\textbf{Key lookup.} The verifier performs the same LPM lookup on the packet's source address using its local key trie. If no matching prefix is found, the verifier follows the local policy and either forwards or drops the packet.\\
\textbf{Signature recomputation.} Using the retrieved key, the verifier computes a SHA-256 hash
over the same header fields. \\
\textbf{Verification and forwarding decision.} The verifier compares the recomputed signature with the one extracted from \texttt{IP Options}. A match means the packet is forwarded. A mismatch means it is processed for invalided packets; e.g. dropped or logged.

\subsection{Prefix Lookup}

Both the sender and the verifier select the authentication key with a Longest Prefix Match over source IP prefixes. This lookup runs on every authenticated packet, so it directly affects latency, memory use, and update cost in all three target environments. For \liha{}, the lookup structure is thus a critical design choice.

\liha{} requires four things from this structure. First, it must support LPM, since keys are bound to IPv4 prefixes of different lengths. Second, it must provide low and predictable per-packet latency, because the lookup is on the critical path. Third, it must support inserts and deletes efficiently, since keys rotate over time and new prefixes may be added incrementally. Fourth, it must use memory conservatively so it can fit in constrained settings such as FPGA on-chip memory and eBPF maps, not just in userspace.

We evaluated five LPM designs: a binary radix trie, a Patricia trie, DIR-24-8, DXR, and DXR with a Bloom filter. DIR-24-8 gave the fastest average lookup, around 10--15 ns. However, it used about an order of magnitude more memory than trie-based designs and made updates more complex. The radix trie and Patricia trie were slower, in the 65--80 ns range, but used much less memory and supported simple local updates. The DXR-based designs fell between these two points.
\liha{} uses a binary radix trie as its lookup structure. This choice is based on overall system fit, not just raw lookup speed.

First, the radix trie maps well to all three target environments. The same basic design can be implemented in FPGA block RAM, in eBPF under verifier limits, and in pointer-based userspace code. Other designs, such as DIR-24-8 and DXR, rely more heavily on large table layouts. Those layouts fit less naturally within FPGA memory limits and are harder to express cleanly under eBPF constraints.

Second, the radix trie works well with \liha{}'s key rotation model. \liha{} does not use sequence numbers for replay protection. Instead, it relies on frequent key rotation. That means inserts and deletes happen regularly and must stay cheap. In a radix trie, updates are local and run in O(L). DIR-24-8, in contrast, requires coordinated changes across multiple tables, which makes frequent updates less attractive.

Third, radix trie latency is already small compared with the rest of the authentication pipeline. In all three environments, the main cost comes from SHA256 and the surrounding packet-processing steps, such as parsing, canonicalization, checksum repair, and reinjection. For example, in eBPF, SHA256 takes about 1{,}300 ns per packet. A radix trie lookup at roughly 100 ns is only a small part of that total. A faster lookup structure would save only tens of nanoseconds while using much more memory.

Fourth, simplicity matters for security. The radix trie is compact and easy to reason about. It can fit within eBPF's 1M-instruction limit and can be synthesized on FPGA without dominating routing or memory resources. Because \liha{} depends on this structure for its trust decision, a simpler design also reduces implementation risk.

In practice, radix trie performance is more than sufficient. Under mixed read/write workloads, it reaches a stable operating point at a read-to-write ratio of 1:2{,}000, with average latency leveling off around 94 ns/op. Real web traffic is well above this threshold. A median 2.9 MB desktop page generates roughly 2{,}000 packets per key-update event, which places normal traffic comfortably within the sustainable range.

More aggressive structures such as DIR-24-8 or DXR could still be good choices in deployments with abundant memory and infrequent key updates. For \liha{}'s target setting, however, the binary radix trie provides the best overall balance.

\section{Keyserver}
\label{sec:keyserver}

\liha{}'s design requires a keyserver. 
Our implementation uses a single HashiCorp Vault server~\cite{sheri2025hashicorp}. Nodes access it through a TLS-secured REST API with mutual certificate authentication. Keys rotate on a configurable schedule. In our evaluation, they rotate every 24\,hours. The keys are stored in a versioned key value store indexed by IP prefix. At startup, and again at each rotation, a node fetches the key once and loads it into its radix trie.

Our approach deliberately moves security issues such as key distribution, and its security, out of the network layer,
unlike almost all prior approaches. Separating the key management layer is actually a strength from design
perspective because it enables a richer and diverse space of key management. Keyservers can scale from simple web-servers to complex, distributed designs without impacting the network layer. We touch on some of these issue in this section, but a full treatment of key management in the context of network elements performing authentication is beyond the scope of this work, which focuses on the network layer. 

{\bf Identity and Access Management.}
Who is allowed to communicate to whom? From the keyserver's perspective, who is 
authorized to obtain secret keys to different IP prefixes? Such decisions belong in an IAM layer, not in the
network layer. By separating the IAM concerns from the network we improve network authentication performance. 

{\bf Key rotation speed.}
In our local network obtaining a new key takes about 30\,ms end to end. About 22\,ms comes from two sequential TLS handshakes. The rest comes from Vault processing and network delay. After the key is loaded, the server is no longer involved in packet processing. Key rotation speed is one leg of preventing replay attacks; if communicating parties can rotate keys faster than attackers can distribute them such attacks can be mitigated. 

{\bf Malevolent actor key identification.}
A second leg of mitigating attacks is key identification of malevolent actors. Key identification means identifying
traffic originating from compromised hosts and routers and canceling the secret keys. It also means linking
the secret keys to public keys in the keyserver and changing their IAM rules. 

{\bf DDoS attacks on the keyserver.}
DDoS attacks on the keyserver can disrupt communications and extend replay attacks. Mitigations include using
distributed keyservers and putting keyservers behind anycast networks. 

{\bf Compromising the keyserver.}
A compromised keyserver could apply denial of service attacks. Cooperating attacks can also generate
malicious traffic.

\section{Limitations}
\label{sec:limitations}

\liha is intentionally limited and does not claim to solve all network security problems. In this section we describe the limitations of the \liha design.

\liha depends on key secrecy. If a key leaks, anyone can forge packets from that prefix until the key is revoked. Scoping keys to a prefix limits the damage, but the operator still has to manage keys carefully.

\liha also does not give full replay protection on its own. We can add replay protection, but this usually needs extra state, timestamps, or sequence numbers. These are useful in some networks, but we kept them out of the basic forwarding path to keep it simple.

The scope of protection is also narrow. \liha authenticates the packet source but it does not protect packet payload. Applications that require payload integrity and confidentiality must rely on existing end-to-end mechanisms. It also cannot force non-participating networks to filter spoofed traffic. \liha's deployment is incremental, so the benefit appears first in networks where operators choose to enforce authentication for selected prefixes.

In the end, \liha depends on operators who actually want to use it. The hardware path is realistic, SRv6 already carries the packet, and the forwarding algorithm is simple. But operators still need to set up keys, define their policy, and choose where authentication is required.

\section{Implementations}
\label{sec:implementation}
Today's network stacks are vast software constructs that challenge where new protocols can
be placed. However, \liha{}'s simplicity allows a range of places in the networking stack
both the sender authentication and verification can be placed: as a userspace process,
as a kernel-level eBPF program, or inside an network interface card (NIC). Each environment
has different development and performance tradeoffs. In this section we describe the
implementations across those three environments. Section~\ref{sec:challenges} will provide our experience with development within each platform.

\subsection{FPGA Implementation}
\label{sec:fpga}

\begin{figure*}[t!]
    \centering
    \includegraphics[width=0.85\textwidth]{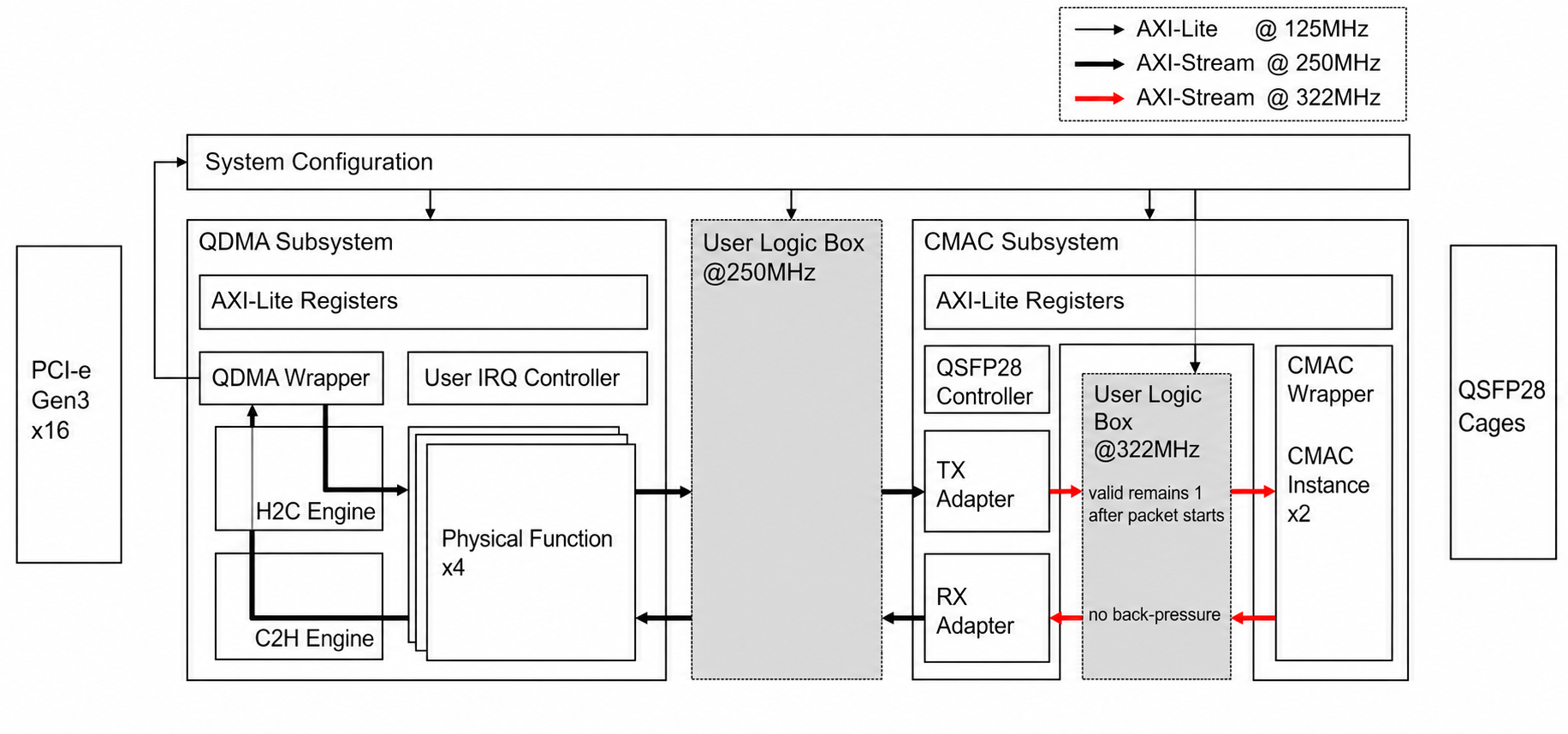}
    \caption{OpenNIC Architecture}
    \label{fig:shell_architecture}
\end{figure*}

The foundation of our implementation is the OpenNIC Shell (Fig.~\ref{fig:shell_architecture}) \cite{d13} for the Xilinx UltraScale+ family of FPGAs. The three main subsystems are: (i)~the QDMA subsystem, which interfaces with the host via PCIe Gen3\,$\times$16 and operates at 250\,MHz; (ii)~the User Logic Region, and (iii)~the CMAC subsystem, which provides access to two 100\,Gbps full-duplex Ethernet ports. A Packet Adapter module bridges the different 250\,MHz PCIe clock domain and the 322\,MHz physical layer clock domain.

\liha{} is realized as Verilog modules added to the central 250 MHz user box in (Fig.~\ref{fig:shell_architecture}) as this module provides back pressure, unlike the 322 MHz box.
All steps are reduced in scope to fit into a 4 ns cycle, and all steps are fully pipelined. Thus, bandwidth is not reduced but adding \liha{} increases latency over the original OpenNIC design. The current prefix tree holds 1000 entries. On both the sending and receiving side, the 64 rounds of the
SHA-256 computation map to one cycle each. Other processing steps, such as header and buffer management require additional cycles; the total added delay through one FPGA averages to 125 cycles.

{\bf FPGA Workflow.} The complete RTL flow is managed inside Vivado, which carries the design from elaboration through synthesis, place--and--route, timing closure, power analysis, and final bit‑stream generation for the Alveo\,U200 FPGA. Functional correctness is first established entirely in software: packets are generated with \textbf{Scapy}, golden SHA‑256 MACs are computed offline in Python, and the design under test (DUT) is driven by coroutine‑based \textbf{Cocotb} test‑benches running under \textbf{ModelSim}, where full waveform visibility accelerates debugging. Full functional simulation completes in under $\mathbf{10}$~minutes, whereas bit‑stream generation and hardware build take on the order of $\mathbf{5}$~hours, making simulation the preferred mode for rapid iteration. Cocotb exposes internal RTL signals, FSM states, and data paths, enabling cycle‑accurate inspection without recompilation. After timing closure, the generated bit‑stream is loaded onto the board and the \emph{same} test suite is replayed through a host‑side driver, providing an end‑to-end hardware check. Vivado’s Integrated Logic Analyzer (ILA) cores are inserted on critical nets so that internal states can be sampled while real traffic flows. This two‑stage \emph{simulate‑then‑silicon} methodology shortens iteration time while ensuring that behaviour validated in software is preserved in deployed hardware.

\subsection{Userspace}
\label{sec:userspace}


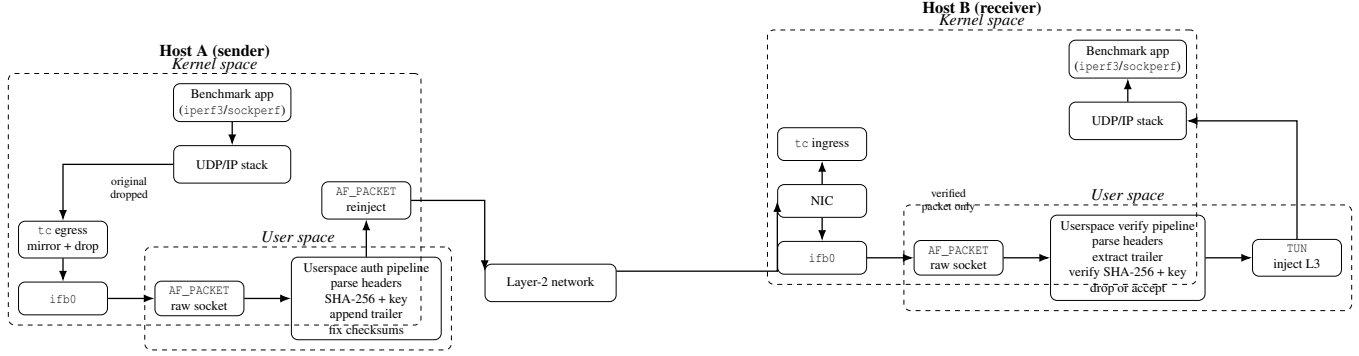
\begin{figure*}[t]
\centering
\resizebox{\textwidth}{!}{%
\begin{tikzpicture}[
    font=\footnotesize,
    >=Latex,
    node distance=5mm and 7mm,
    block/.style={
        draw,
        rounded corners,
        align=center,
        minimum width=2.5cm,
        minimum height=0.8cm
    },
    smallblock/.style={
        draw,
        rounded corners,
        align=center,
        minimum width=1.9cm,
        minimum height=0.7cm
    },
    group/.style={
        draw,
        dashed,
        rounded corners,
        inner sep=6pt
    },
    flow/.style={->, thick}
]

\node[block] (appA) at (0,3.0) {Benchmark app\\(\texttt{iperf3}/\texttt{sockperf})};
\node[block, below=of appA] (udpA) {UDP/IP stack};

\node[smallblock, below left=8mm and 14mm of udpA] (tcA) {\texttt{tc} egress\\mirror + drop};
\node[smallblock, below=of tcA] (ifbA) {\texttt{ifb0}};
\node[smallblock, right=10mm of ifbA] (afA) {\texttt{AF\_PACKET}\\raw socket};

\node[block, right=10mm of afA, minimum width=3.2cm, minimum height=1.5cm] (authA)
{Userspace auth pipeline\\
parse headers\\
SHA-256 + key\\
append trailer\\
fix checksums};

\node[smallblock, above=8mm of authA] (reinjA) {\texttt{AF\_PACKET}\\reinject};

\node[group, fit=(appA)(udpA)(tcA)(ifbA)(reinjA)] (kernelA) {};
\node[group, fit=(afA)(authA)] (userA) {};

\node[font=\bfseries] at ($(kernelA.north)+(0,0.45)$) {Host A (sender)};
\node[font=\itshape] at ($(kernelA.north)+(0,0.18)$) {Kernel space};
\node[font=\itshape] at ($(userA.north)+(0,0.18)$) {User space};

\node[block, minimum width=2.8cm] (net) at (6.8,-0.85) {Layer-2 network};

\node[smallblock] (nicB) at (12.6,0.9) {NIC};
\node[smallblock, above=of nicB] (tcB) {\texttt{tc} ingress};
\node[smallblock, below=of nicB] (ifbB) {\texttt{ifb0}};

\node[smallblock, right=10mm of ifbB] (afB) {\texttt{AF\_PACKET}\\raw socket};

\node[block, right=10mm of afB, minimum width=3.3cm, minimum height=1.5cm] (verB)
{Userspace verify pipeline\\
parse headers\\
extract trailer\\
verify SHA-256 + key\\
drop or accept};

\node[smallblock, right=10mm of verB] (tunB) {\texttt{TUN}\\inject L3};

\node[block, above=16mm of verB] (udpB) {UDP/IP stack};
\node[block, above=of udpB] (appB) {Benchmark app\\(\texttt{iperf3}/\texttt{sockperf})};

\node[group, fit=(nicB)(tcB)(ifbB)(udpB)(appB)] (kernelB) {};
\node[group, fit=(afB)(verB)(tunB)] (userB) {};

\node[font=\bfseries] at ($(kernelB.north)+(0,0.45)$) {Host B (receiver)};
\node[font=\itshape] at ($(kernelB.north)+(0,0.18)$) {Kernel space};
\node[font=\itshape] at ($(userB.north)+(0,0.18)$) {User space};

\draw[flow] (appA) -- (udpA);
\draw[flow] (udpA.west) -| (tcA.north);
\draw[flow] (tcA) -- (ifbA);
\draw[flow] (ifbA) -- (afA);
\draw[flow] (afA) -- (authA);
\draw[flow] (authA) -- (reinjA);
\draw[flow] (reinjA.east) -| ($(net.west)+(0,0.25)$);

\draw[flow] ($(net.east)+(0,0.25)$) -| (nicB.west);

\draw[flow] (nicB) -- (tcB);
\draw[flow] (nicB) -- (ifbB);
\draw[flow] (ifbB) -- (afB);
\draw[flow] (afB) -- (verB);
\draw[flow] (verB) -- (tunB);
\draw[flow] (tunB.north) |- (udpB.east);
\draw[flow] (udpB) -- (appB);

\node[align=center, font=\scriptsize] at (-2.2,1.15) {original\\dropped};
\node[align=center, font=\scriptsize] at (15.3,0.9) {verified\\packet only};

\end{tikzpicture}%
}
\caption{Userspace packet-authentication architecture. On Host A, \texttt{tc} mirrors matched egress traffic to \texttt{ifb0}, drops the original packet, and hands the frame to userspace through an \texttt{AF\_PACKET} socket for authentication and reinjection. On Host B, intercepted ingress traffic is verified in userspace; valid packets are delivered through a \texttt{TUN} interface and invalid packets are dropped.}
\label{fig:cpu-userspace-arch}
\end{figure*}

This section outlines the userspace architecture employed in our experimental setup to evaluate CPU-bound packet authentication overhead in a controlled environment. The objective is to assess the practicality and cost of cryptographic packet verification and what fraction of the line rate is obtainable without specialized network interface or additions to the kernel. 

\subsubsection{Architecture}

The userspace implementation evaluates packet authentication in a live userspace pipeline using standard Ethernet/IP traffic. The experimental setup consists of two Linux hosts connected within the same Layer-2 domain, with benchmarking traffic generated using \texttt{iperf3} for bandwidth measurements and \texttt{sockperf} for round-trip latency measurements. This design isolates the cost of userspace packet authentication under an application-visible workload while avoiding protocol-specific artifacts unrelated to the authentication pipeline itself.

The architecture moves packet authentication into userspace using \texttt{AF\_PACKET} sockets together with Linux traffic-control primitives for interception and enforcement. This avoids the verifier and instruction-limit constraints associated with eBPF, but introduces a different set of systems challenges involving packet interception, egress enforcement, reinjection safety, and bidirectional delivery correctness.

\subsubsection{Packet Interception}

Live experiments require interception at a point where the kernel can be instructed to either drop, accept, or redirect traffic before it continues along the normal path. For the Ethernet/IP pipeline used in the \texttt{iperf3} throughput and \texttt{sockperf} RTT experiments, interception is implemented using Linux traffic control (\texttt{tc}) with mirroring to an Intermediate Functional Block (IFB) device together with drop enforcement on the original path.

The IFB acts as an explicit handoff boundary between kernel forwarding and userspace processing. Packets matching the configured classifier are copied to \texttt{ifb0}, while the original packet is prevented from continuing along the normal transmit path. Userspace then reads complete Layer-2 frames from \texttt{ifb0} through an \texttt{AF\_PACKET} raw socket and applies authentication logic before reinjection. This design provides both enforcement and controlled delivery of the in-scope traffic class to userspace without racing the physical NIC transmit queue.

Correctness depends on two requirements. First, enforcement must be strict: if the original packet is still transmitted while userspace processes a copy, the system either produces duplicates or allows unauthenticated traffic to bypass authentication whenever userspace falls behind. Second, userspace must observe a faithful representation of the packet that would have been sent or received, since authentication depends on consistent parsing and canonicalization of the network and transport headers.

\subsubsection{Packet Modification}

Packet modification converts an in-scope UDP packet into an authenticated packet. In the live Ethernet/IP pipeline, authentication metadata is appended as a compact trailer to the end of the UDP payload rather than inserted into the IPv4 options field. This keeps the header format stable while still carrying authentication material in-band.

The sender parses the Ethernet header, including optional VLAN tags, then locates the IPv4 and UDP headers and confirms that the packet belongs to the benchmark flow. Authentication is computed using SHA-256 over a canonicalized representation of the packet headers together with secret key material. Canonicalization is required so that sender and receiver compute the same digest even when some on-wire fields are updated during forwarding or recomputed by the kernel. After digest computation, the sender appends the authentication trailer, updates the IPv4 and UDP length fields, and recomputes the corresponding checksums.

This modification path is designed so that verification remains deterministic while the resulting packet remains valid on the wire. The dominant cost is therefore not limited to the hash function itself. At higher packet rates, overhead also includes parsing, canonicalization, metadata insertion, memory movement, checksum repair, and the surrounding userspace diversion mechanics.

\subsubsection{Packet Reinjection}

Packet reinjection restores forward progress after in-line interception. On egress, the authenticated packet is reinjected by transmitting a complete Ethernet frame through an \texttt{AF\_PACKET} raw socket bound to the intended transmit interface. Because the intercepted buffer already contains the original Ethernet header, the reinjection path can reuse the destination MAC address directly and does not require ARP resolution or Ethernet header synthesis. The transmitted frame is therefore identical to the original packet except for the authentication-related modifications introduced during packet modification.

On ingress, reinjection is fundamentally different. Sending through \texttt{AF\_PACKET} transmits a frame out of an interface, but does not deliver it upward as a locally received packet. Verified inbound traffic is therefore injected into the host networking stack through a TUN interface. After verification succeeds, userspace extracts the Layer-3 packet and writes the IP packet bytes to the TUN file descriptor, allowing the kernel to treat the packet as locally received traffic and deliver it through the normal transport stack to the application.

Loop safety is also required once reinjection is introduced. Reinjected packets must not be captured again by the same interception rules. This is handled through stable packet markers together with classifier exclusions, so that authenticated packets bypass repeated interception rather than re-entering the IFB path.



The user-space implementation is a daemon that embodies the most accessible \liha{} realization.
It captures packets sent by any user process from the Linux kernel and reinjects modified
ones back. It intercepts incoming packets to verify the source signature before they reach the
destination process. 

Linux \texttt{tc} is used to mirror in-scope traffic to an IFB device and to enforce dropping on the original path. The userspace implementation reads complete Layer-2 frames from \texttt{ifb0} through an \texttt{AF\_PACKET} raw socket. This interception method provides a stable handoff point between kernel forwarding and userspace processing and prevents unauthenticated packets from continuing on the normal transmit path.

For each in-scope UDP packet, the  \liha{} daemon parses the Ethernet, IPv4, and UDP headers, canonicalizes the packet representation, and computes a SHA-256 digest using secret key material. Authentication metadata is appended as a compact trailer to the UDP payload. The IPv4 and UDP length fields are then updated, and the corresponding checksums are recomputed.


Authenticated outbound packets are reinjected as complete Ethernet frames through an \texttt{AF\_PACKET} socket. Verified inbound packets are delivered into the local host stack through a TUN interface. Classifier exclusions and packet markers are used to prevent duplicate delivery, bypass, and reinjection loops. The measured overhead therefore includes not only SHA-256 computation, but also interception, parsing, canonicalization, trailer insertion, checksum repair, reinjection, and kernel-userspace boundary crossings.

\subsection{eBPF}
\label{sec:ebpf} 

\begin{figure}[h!]
     \centering
 \includegraphics[width=0.5\textwidth]{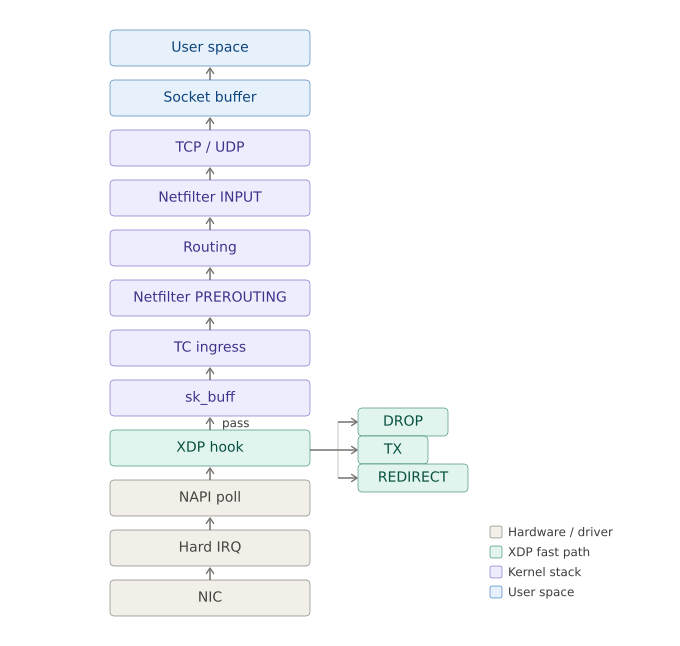}
     \caption{eBPF Architecture}
     \label{fig:eBPF_arch}
 \end{figure}

eBPF is an emerging technology enabling safe packet manipulation inside the kernel's address space.
Recall eBPF programs are written in systems programming languages such as C or Rust, and
compiled into an eBPF specific instruction set. Programs are validated as safe by a verifier
before being allowed to run in the kernel; loading them into the kernel incurs an addition
step of translating the eBPF virtual machine instructions into native ones. 

\liha{}'s simplicity allows for realizing most of it inside of a pair of eBPF programs;
the ingress traffic uses the eBPF eXpress Data Path (XDP) hook, and egress uses the
Traffic Control (TC) hook. The hooks are points in network stack for packet manipulation. 

This section details the architecture of the eBPF based source authentication scheme.
The design implements a custom SHA-256 hash of the IP header \cite{sha256-custom2} in eBPF that is inserted into \texttt{IP Options} for verification on the receiver side.


As shown in Figure~\ref{fig:eBPF_arch} XDP (eXpress Data Path) is a kernel framework that 
provides the earliest possible hook point for packet processing in the Linux networking stack at the Network driver level, before socket buffer (SKB) allocation. However it is available only for ingress traffic. On the other hand, TC (Traffic Control) is a comprehensive kernel subsystem for traffic management, classification, queuing, and shaping that provides hooks after SKB allocation and is available for both ingress and egress traffic.


eBPF is designed for simple, short, programs as the kernel is not made for long running code
\cite{miano2023fast}. 
The verifier restrictions keep the program size to 1M dynamic instructions; that is, programs
that would run for more than 1M instructions are rejected. As an alternative, the SHA-256 code was offloaded to a kernel module and exposed to eBPF as a kernel function (kfunc)~\cite{sharaf2022extended}. These are trusted functions that run without eBPF limits and enable unparalleled development agility. They allow developers to write their own bespoke kfuncs instead of relying on limited BPF helper functions. As another alternative, the Linux kernel crypto APIs were used instead of the custom SHA-256 code in a kernel module. The Linux kernel 6.10+ also exposes crypto APIs as kfuncs that can be used by eBPF. An advantage of the Crypto API is that this library
uses hardware acceleration if available. CPUs with Intel SHA extensions have demonstrated a 3.6x speedup over optimized SSSE3 software routines.

We found \liha{} easily fits as an eBPF program when SHA-256 is a kfunc, and can fit many
systems where the SHA-256 must run as a full eBPF program. eBPF CO-RE feature allows developers to write a single eBPF program that runs across multiple Linux kernel versions without recompilation. The prefix lookups use the
lookup code described in Section~\ref{sec:cpu-lookup}.

\section{Experiment Results}
\label{sec:results}

\begin{figure}[th]
    \centering
    \includegraphics[width=0.48\textwidth]{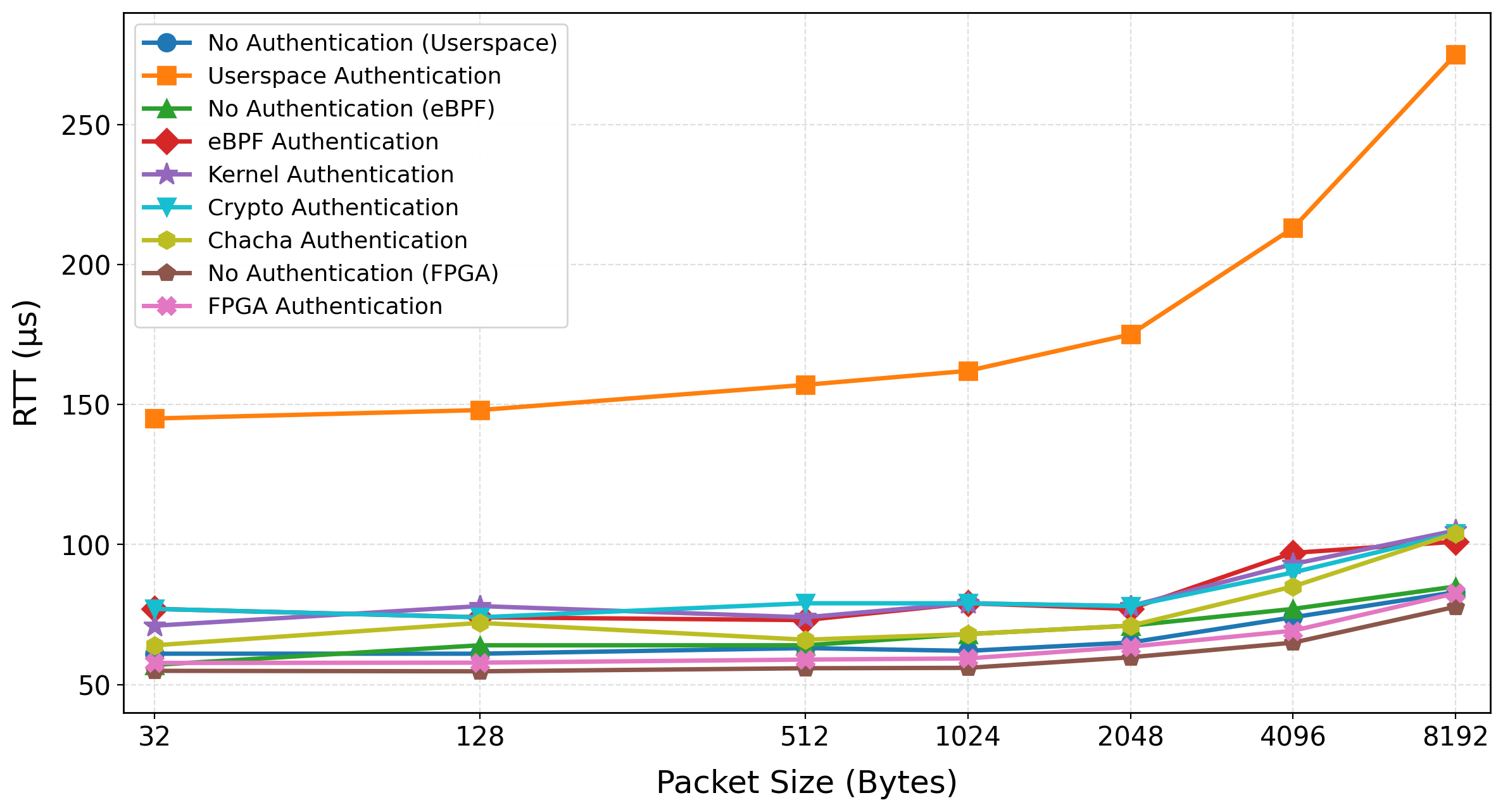}
    \caption{Round-Trip Time Performance for all platforms.}
    \label{fig:all-latency}
\end{figure}

\begin{figure*}[tp]
    \centering
    \includegraphics[width=0.8\textwidth]{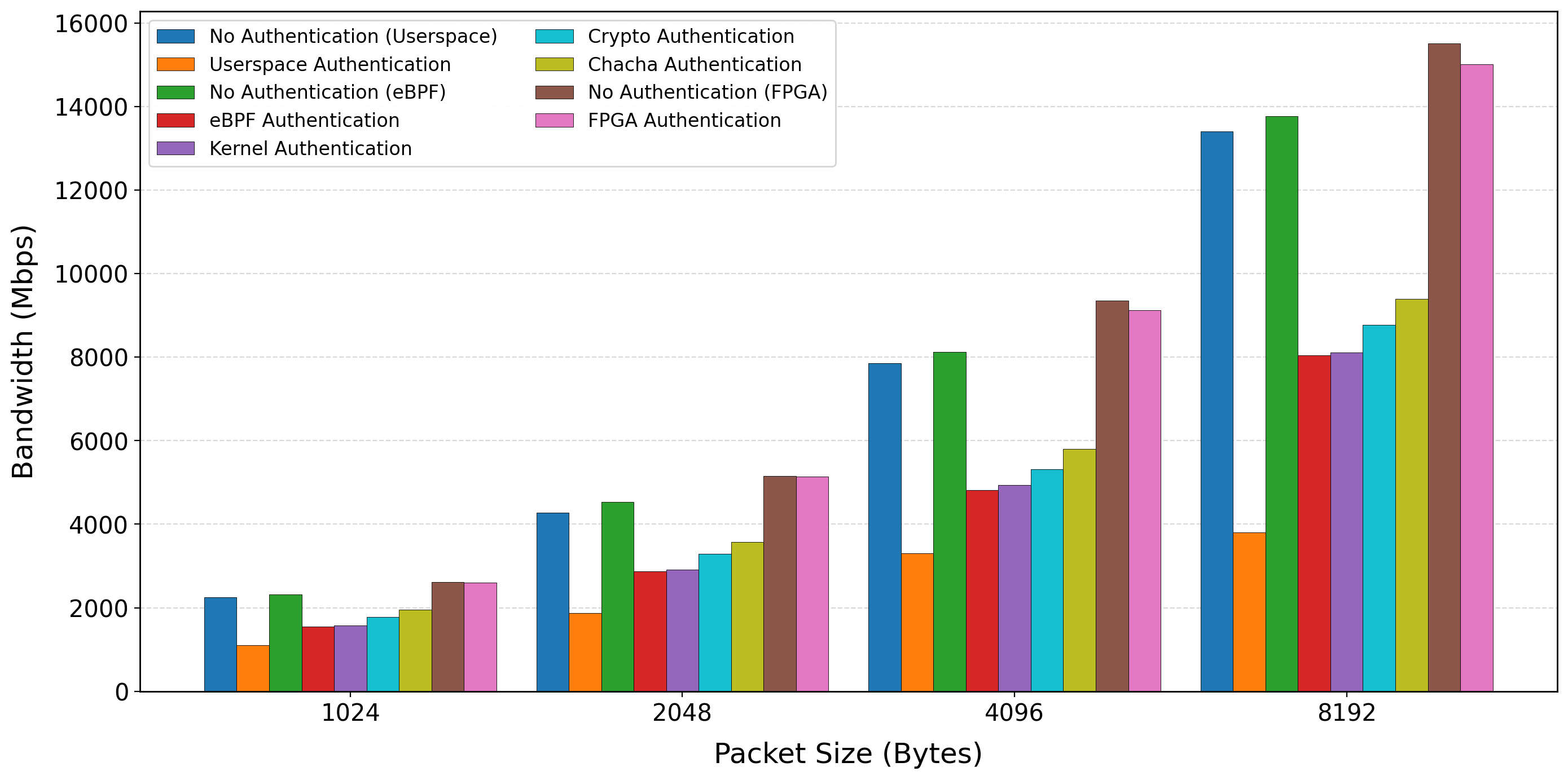}
    \caption{Bandwidth Performance for all platforms.}
    \label{fig:all-throughput}
\end{figure*}

In this section we present two key results. First, we show \liha{}'s incremental cost over
standard UDP is low. Space limitations do not allow for a full performance characterization, but we summarize
\liha{}'s incremental costs by first showing prefix lookup costs, and then
use UDP micro-benchmarks over the three implementations.
Second, we show it's effectiveness at DDoS mitigation using an emulated attack that scales to 100 Gbps.

\subsection{Prefix Lookup}
\label{sec:cpu-lookup}

\begin{figure*}[t]
    \centering
    \begin{tabular*}{\textwidth}{@{\extracolsep{\fill}}ccc}
        \includegraphics[width=0.32\textwidth]{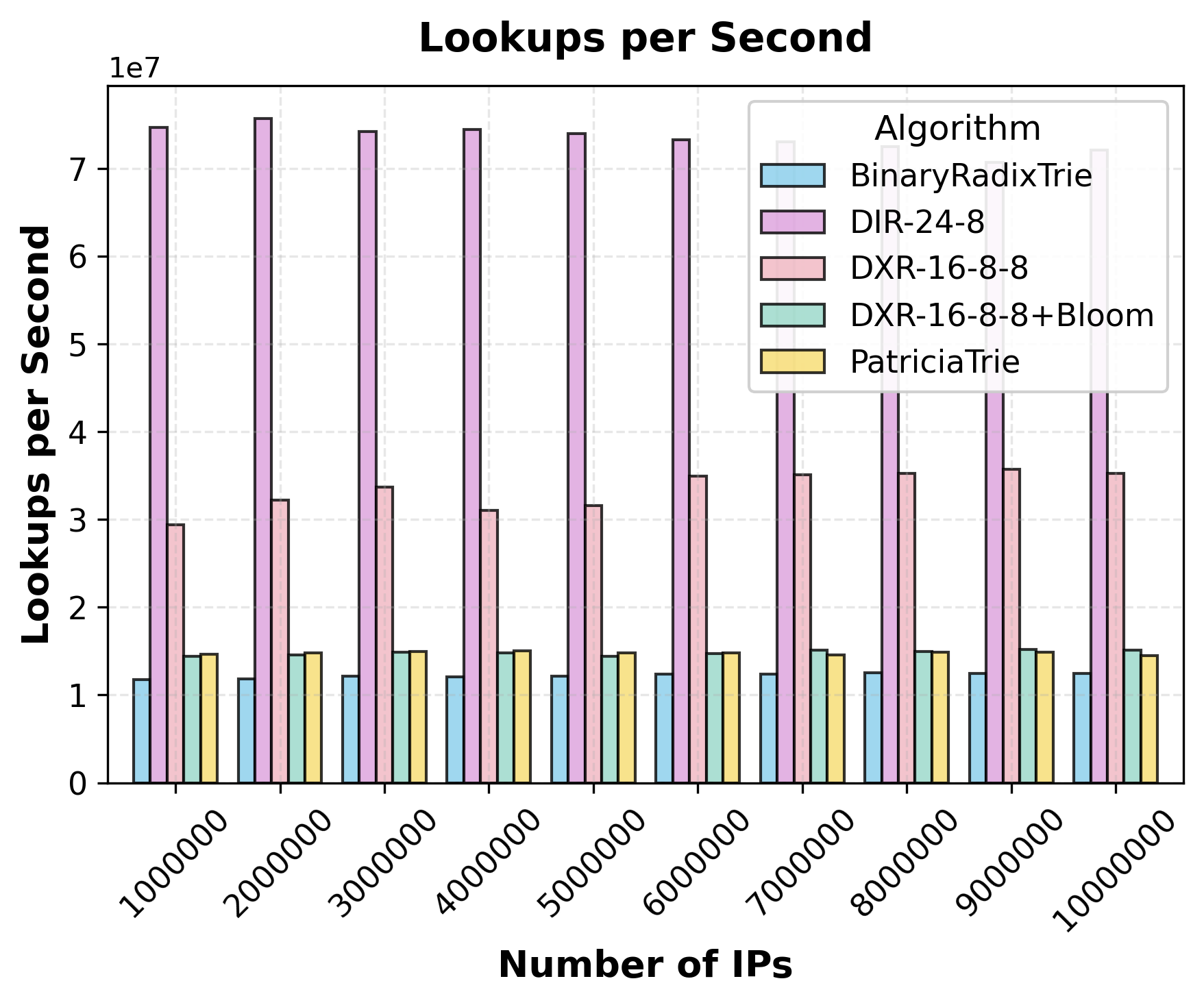} &
        \includegraphics[width=0.32\textwidth]{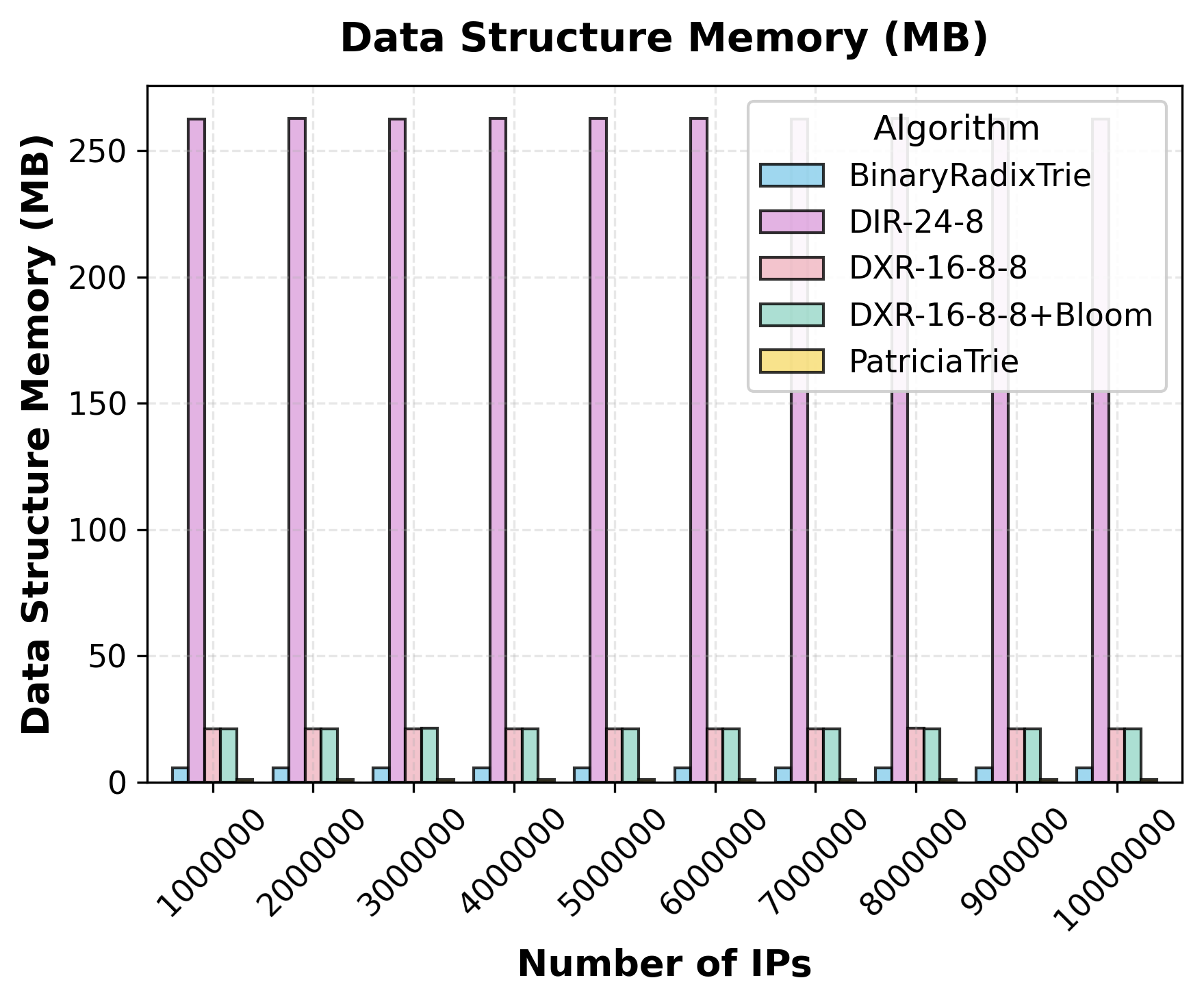} &
        \includegraphics[width=0.32\textwidth]{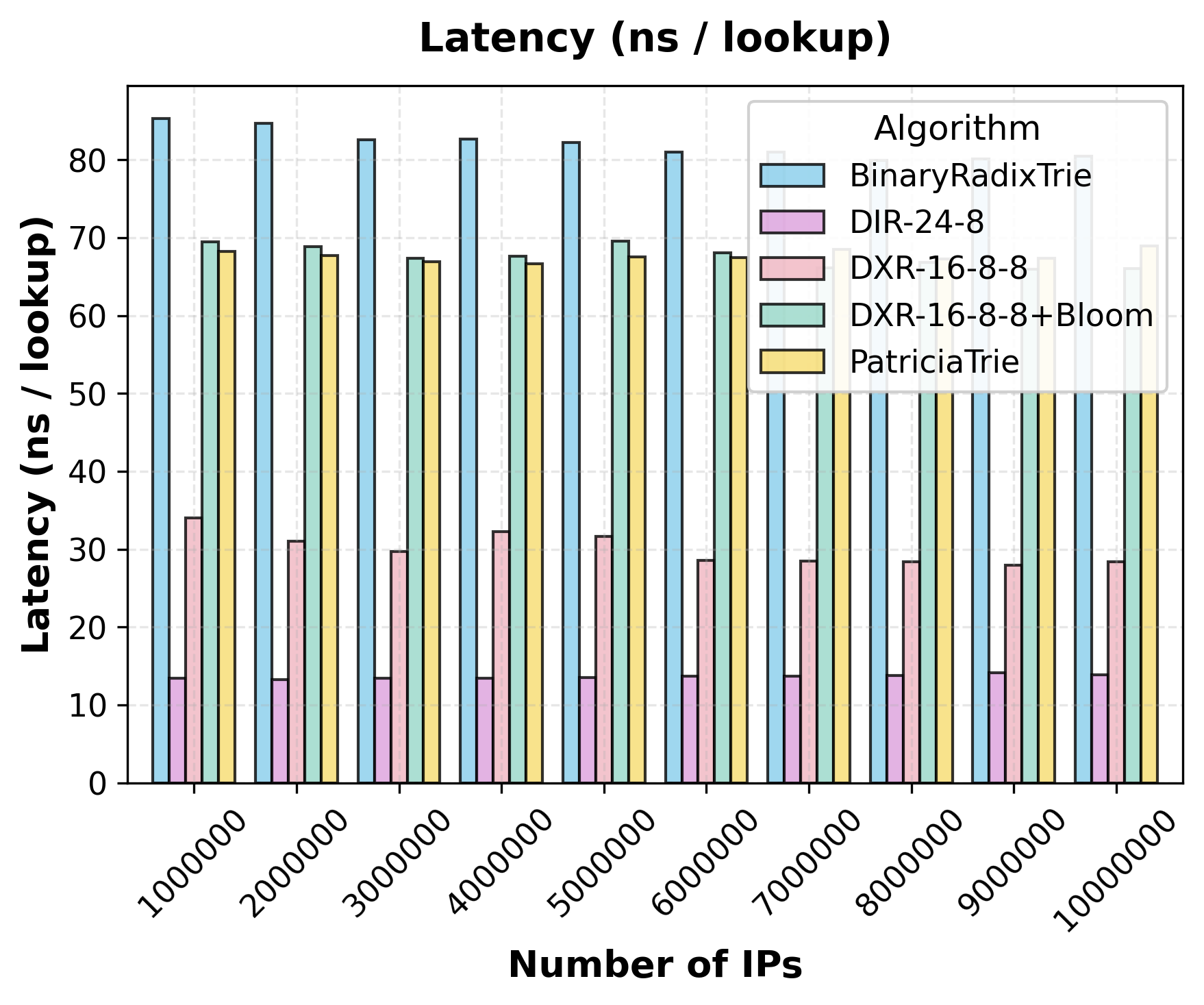} \\
        (A) & (B) & (C) \\
    \end{tabular*}
    \caption{Lookup performance evaluation: (A) Lookup throughput across increasing lookup counts; (B) Memory usage of each lookup structure; (C) Average lookup latency in nanoseconds.}
    \label{fig:combined-lookup-results}
\end{figure*}

Longest Prefix Match (LPM) has long been used IP prefixes as keys to lookup values, for example, a
destination port in a router \cite{sklower1991tree,waldvogel1997scalable}. 
\liha{} leverages the existing prefix lookup machinery although the returned value is
a 64 byte secret key rather than a destination port. As \liha{}'s first step requires a prefix match, 
we implemented and evaluated five IP lookup algorithms: Binary Radix Trie, Patricia Trie, DIR-24-8, DXR, and DXR with Bloom filters (DXR Bloom). The goal was to compare their lookup latency, throughput, memory usage, and build time under both static and dynamic workloads. 


\paragraph{Binary Radix Trie}
A straightforward bitwise trie that moves one bit per level \cite{radix_trie_reference}. It is easy to implement and its behavior is predictable, but it uses a large amount of memory and suffers from poor cache locality. Updates take $O(L)$ time, where $L$ is the prefix length.

\paragraph{Patricia Trie}
A compressed version of the radix trie that merges long chains of single-child nodes \cite{patricia1968}. This reduces pointer depth and memory usage while improving cache performance. Insertions and deletions require structural changes when prefixes need to be split or merged.

\paragraph{DIR-24-8}
A two-level lookup table that directly indexes the upper 24 bits of an address \cite{gupta1998routing}. Entries either point to a result or to a small secondary table indexed by the lower 8 bits. Lookups need no more than two memory accesses and are close to constant time. Dynamic updates are coordinated through a helper radix trie to maintain consistency across the tables.

\paragraph{DXR}
The Dynamic Exact Range method replaces subtables with compact vectors of prefix ranges \cite{dxr2004}. Lookups perform a binary search over these ranges, reducing memory usage and improving cache behavior compared to DIR-24-8.

\paragraph{DXR Bloom}
DXR Bloom adds a Bloom filter \cite{bloom1970} to quickly check for prefixes that are not present, allowing the lookup to skip unnecessary searches. This is a custom design combining DXR \cite{dxr2004} with Bloom filtering. Our test data was constructed so that every lookup address always has a matching prefix, which means the Bloom filter never returns negative. In this case, it only adds a small constant overhead due to computing the hash values. We include DXR Bloom to show that probabilistic filtering does not significantly degrade performance even in a worst-case scenario where it provides no benefit.

\paragraph{Measurement Setup}
All Longest Prefix Match (LPM) algorithms were evaluated using a single-threaded implementation, executing sequentially on an Intel Core i7-12700H processor (12th Generation Alder Lake architecture) with a maximum turbo frequency of 4.7~GHz. The CPU features 14 physical cores (composed of a hybrid performance/efficiency core layout) and a 24~MB shared L3 cache with a 64-byte cache line alignment. To evaluate real-world production performance, all custom implementations were compiled using the \texttt{-O2} optimization flag, allowing the compiler to perform instruction scheduling and global optimization. No explicit thread affinitization or core pinning was enforced, leaving thread scheduling entirely to the native operating system scheduler. All benchmarks were executed on a 64-bit Kali GNU/Linux (Rolling Edition, version 2026.1) operating system.

All experiments used synthetic FIBs containing 10,000 to 100,000 prefixes and up to 10 million lookups per run. For each algorithm we measured: (1) average lookup time in nanoseconds, (2) total lookup throughput in lookups per second, (3) memory usage in MB, and (4) build time in seconds. Figure~\ref{fig:combined-lookup-results} shows these results.

\paragraph{Mixed Read/Write Workloads and Sustainable Load}
Real systems will need to update the trie as new prefixes are added and old ones removed. 
We evaluated the radix trie under mixed workloads with varying read-to-write ratios, simulating realistic routing scenarios where lookups dominate updates. We measured average operation latency across ratios from 1:1 (equal reads and writes) to 1:1,000,000 (nearly read-only). To identify the sustainable load threshold where write operations no longer significantly impact performance, we applied a plateau detection algorithm that finds where the relative change in average latency becomes consistently small (less than 3\% over a rolling window). The plateau begins at a read-to-write ratio of 1:2,000, corresponding to an average latency of approximately 94~ns/op. This indicates that at ratios of 1:2,000 or higher, the system reaches a sustainable operating point where write operations have minimal impact on overall throughput.

\begin{figure}[t]
    \centering
    \includegraphics[width=0.95\linewidth]{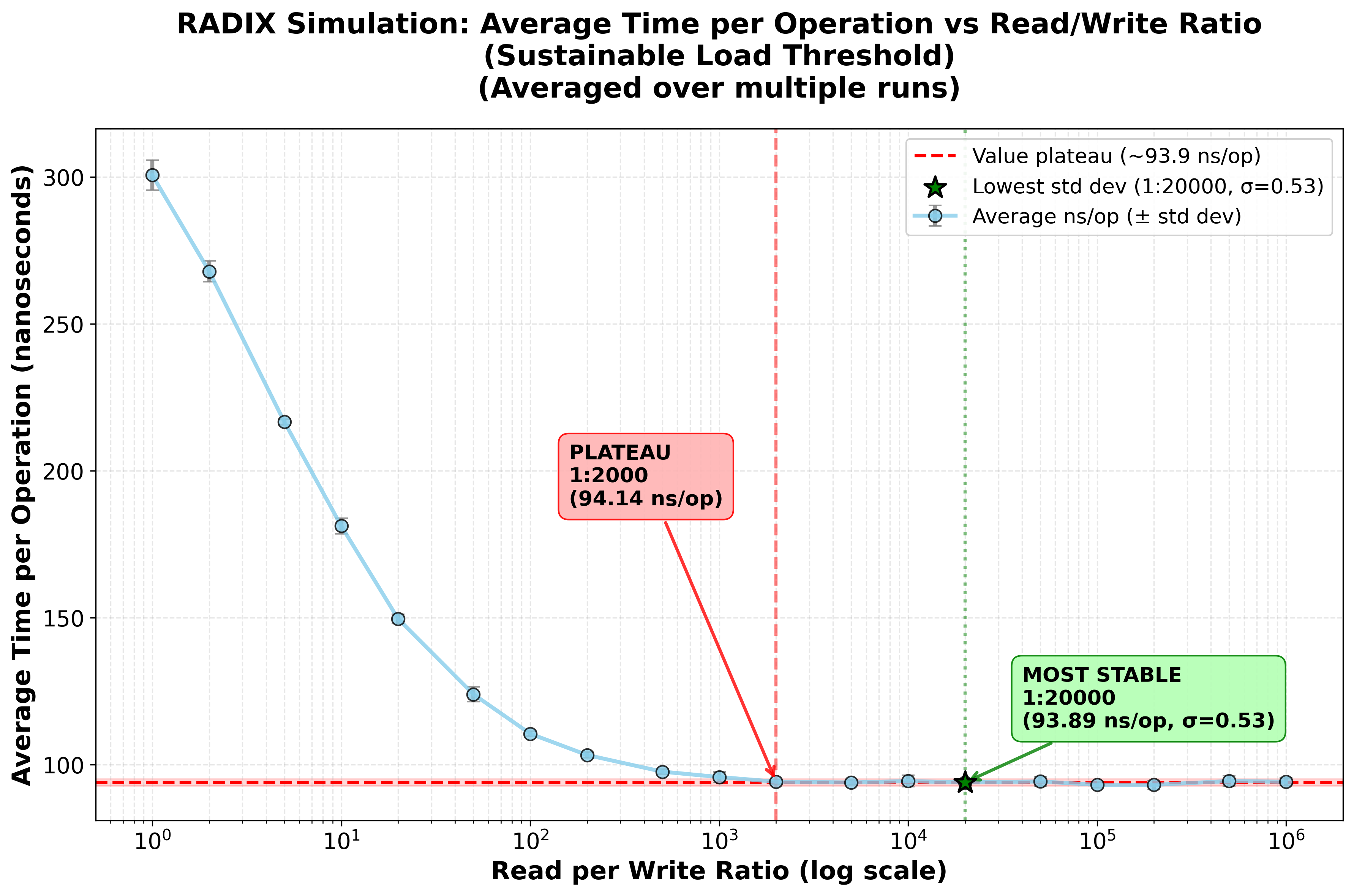}
    \vspace{-1em}
    \caption{Average operation latency across varying read-to-write ratios. The horizontal dotted line indicates the sustainable load threshold at ratio 1:2,000.}
    \label{fig:sim-radix}
\end{figure}

\paragraph{Choosing Radix Trie}
The algorithms exhibit a clear time-versus-memory tradeoff, and thus the choice of algorithm depends on system constraints. Hash table-based approaches like DIR-24-8 offer fast lookups but require significantly more memory. Trie-based approaches such as the binary radix trie and Patricia trie are slower but consume less space and are more amenable to dynamic inserts and deletes, unlike hash tables which may require rehashing or sorting after modifications.

Given our evaluation results, all our implementations use the radix trie because it is the simplest implementation among the approaches. It has the best overall balance between implementation simplicity, lookup speed, memory
footprint, and time to insert and delete entries. 

\begin{figure*}[t]
    \centering
    \begin{tabular}{cc}
        \includegraphics[width=0.35\textwidth]{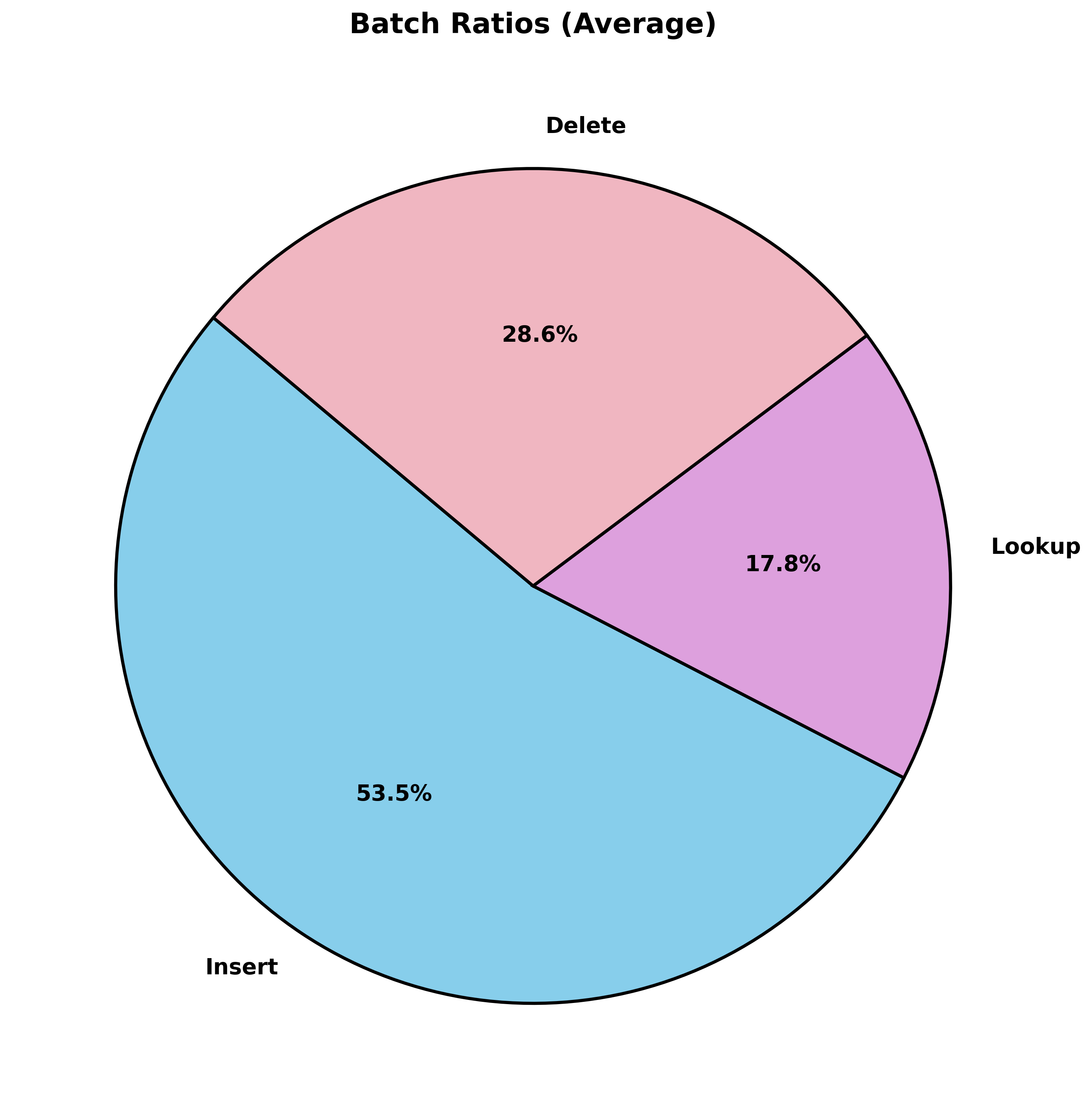} & 
        \includegraphics[width=0.35\textwidth]{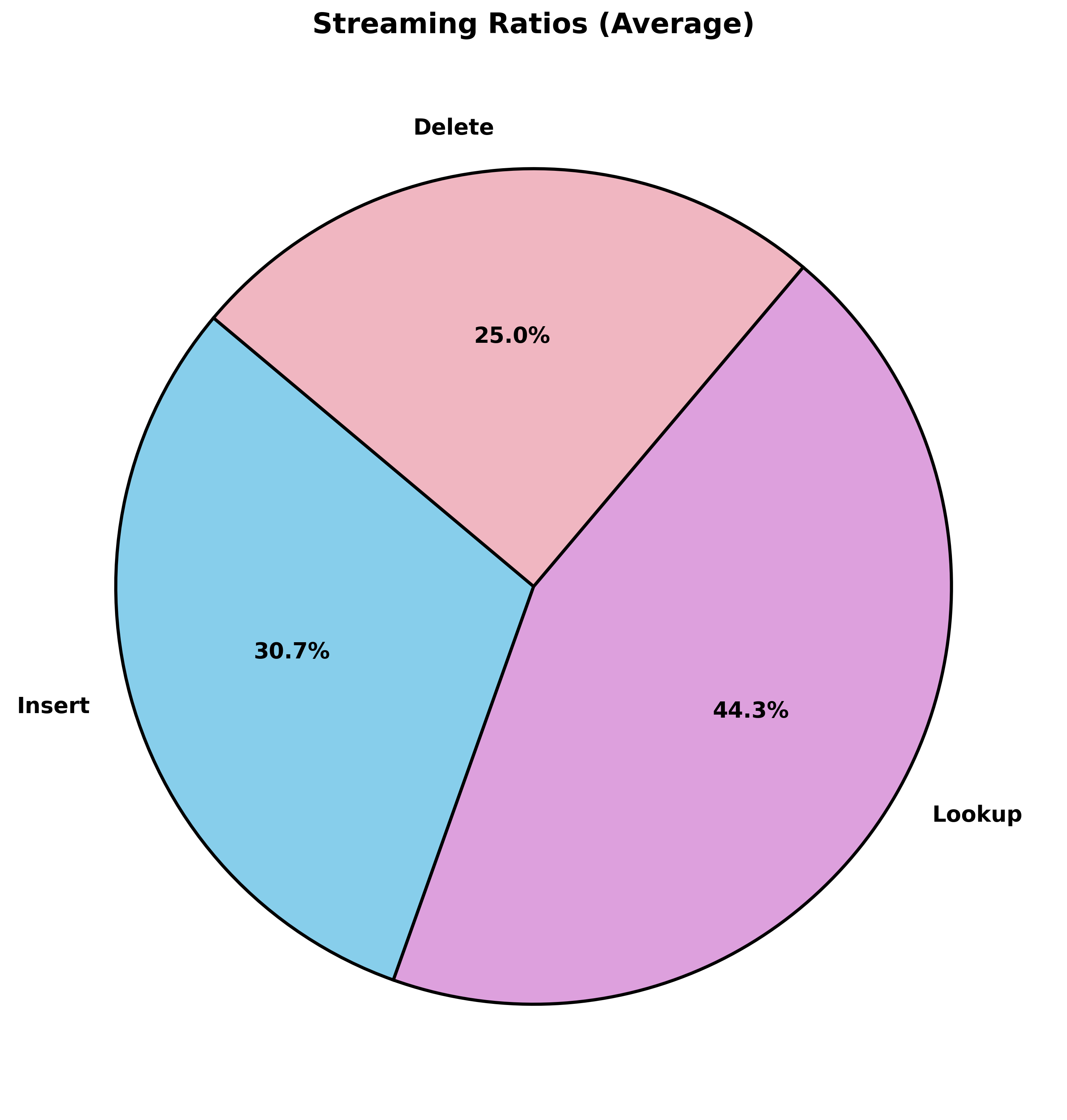} \\
        \textbf{(A)} Batch operation ratios & \textbf{(B)} Streaming operation ratios \\
    \end{tabular}
    \caption{Operation distribution under batch and streaming workloads.}
    \label{fig:ops-radix-pie}
  \end{figure*}

The cost of insert operations dominates when considering dynamic updates to the data structures. This is especially true for DIR-24-8, which would require reorganizing table entries after every insert to maintain consistency.
Lookup operations utilize cache locality better than inserts and deletes. This is evident from Figure~\ref{fig:ops-radix-pie}, where the percentage of time spent on lookups shrinks when operations are batched compared to streaming mode. Batched lookups benefit from warmer caches since consecutive lookups access similar memory regions.

 In a mixed workload simulation where inserts and deletes are interleaved with lookups, the average cost per operation stabilizes as we increase the number of reads performed per write. Beyond a ratio of 1:2{,}000 reads per write, the system reaches a sustainable operating point where write overhead becomes negligible, achieving consistent throughput suitable for production routing workloads. This operating regime is representative of real-world web traffic characteristics: according to HTTP Archive measurements, the median page weight for desktop websites during October--December 2025 is approximately 2.9~MB and continues to increase over time~\cite{httparchive-pageweight}. Assuming a standard Ethernet MTU of 1500~bytes and an effective TCP payload of approximately 1460~bytes per packet, serving a single median desktop page entails on the order of $2.9~\text{MB} / 1460 \approx 2{,}000$ packet-level forwarding reads. Consequently, realistic traffic naturally operates beyond the 1:2{,}000 read-to-write threshold, indicating that radix-trie--based forwarding structures remain scalable under production workloads.

\begin{figure}[t]
    \centering
    \includegraphics[width=0.5\textwidth]{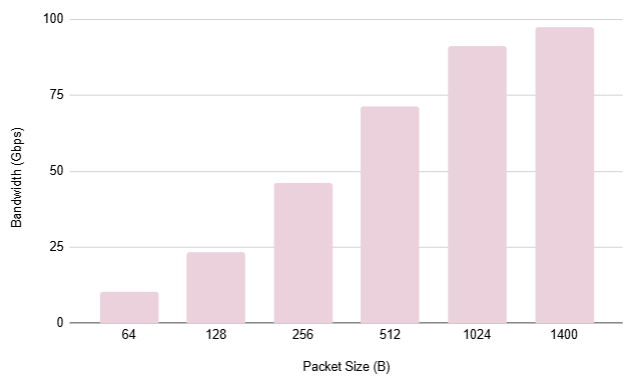}
    \caption{Bandwidth Performance using DPDK. }
    \label{fig:DPDK-throughput}
\end{figure}

\begin{figure}[t]
    \centering
    \includegraphics[width=0.5\textwidth]{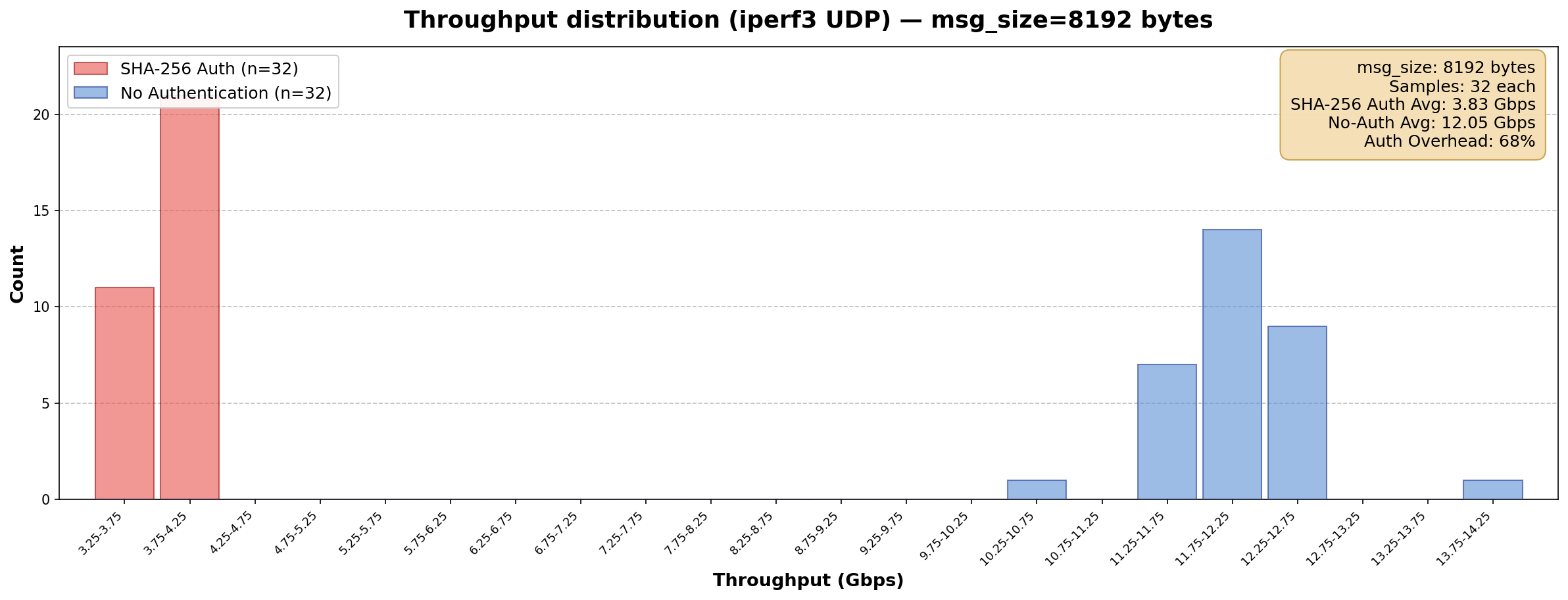}
    \caption{Bandwidth Distribution Userspace (8192B). }
    \label{fig:cpu-throughput-8k}
\end{figure}

\begin{figure}[t]
    \centering
    \includegraphics[width=0.5\textwidth]{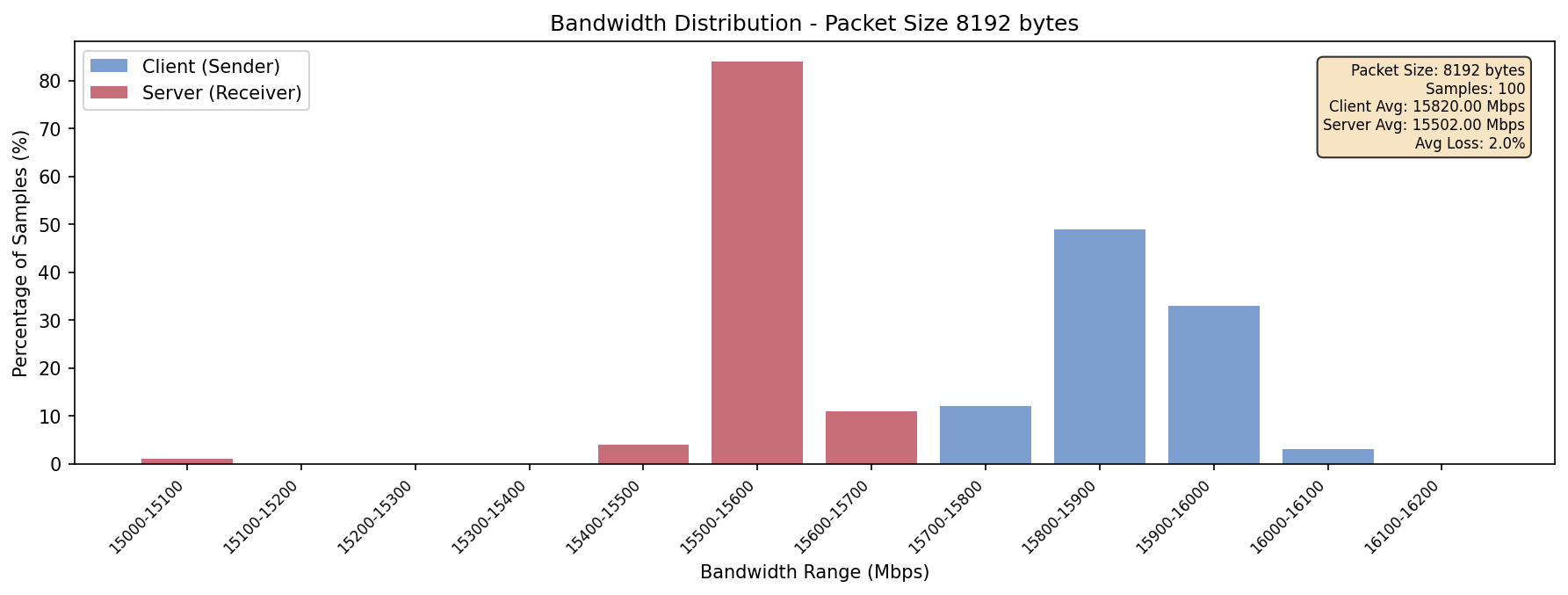}
    \caption{Bandwidth Distribution FPGA (8192B). }
    \label{fig:fpga-throughput-8k}
\end{figure}

\begin{figure}[t]
    \centering
    \includegraphics[width=0.5\textwidth]{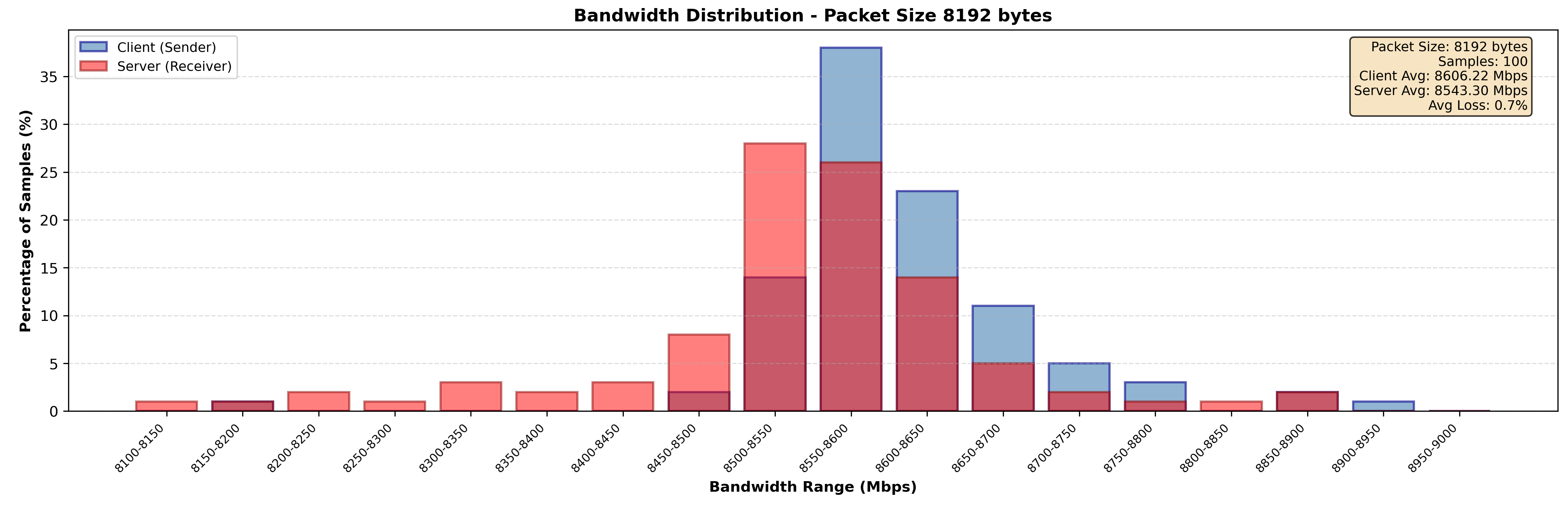}
    \caption{Bandwidth Distribution eBPF (8192B). }
    \label{fig:ebpf-throughput-8k}
\end{figure}

\subsection{Platform Comparison}
In this section we examine latency and bandwidths costs.
In these experiments, all machines were Dell R740XD servers with 192GB of RAM, 256GB SSD.
The installed CPU for these servers was an Intel Xeon Gold 6126 at 2.60GHz.
A 100Gb Mellanox ConnectX-5 was the primary Ethernet, which was used for the Userspace and eBPF experiments.
The FPGA experiments used Xilinx Alveo U200 cards running OpenNIC. All the NICs were connected to a Dell
Z9332F-ON 100Gbps per-port switch. FPGA experiments were run on Ubuntu 22.04 and all others on Ubuntu 24.04.

Figure~\ref{fig:all-latency} shows the measured average round-trip time (RTT) for over 1000 packets
from two servers for a baseline of standard networking
and the \liha{} implementations. The servers were connected by the Dell switch, so the RTT includes the
time through it. The eBPF and FPGA platforms have the lowest latency, with eBPF edging out the FPGA for
very small packets.The Userspace implementation adds significant latency, as the many packet captures, kernel
crossings and packet injections take their toll.

Overall, for wide-area and and many data-center type applications, adding \liha{} using eBPF adds a tolerable
amount of latency; even for large packets the RTT only increases by about 12 $\mu$sec. The FPGA increases
the RTT by only 2-4 $\mu$sec for all packet sizes. The userspace implementation adds significant overheads.

These experiments ran the iperf network measurement tool running UDP and measured receiver bandwidth
~\cite{dugan2010iperf}. 

The bandwidth distributions at 8K packet sizes, that is the fraction of times a given bandwidth
is measured over an interval, (\ref{fig:cpu-throughput-8k}, \ref{fig:fpga-throughput-8k} and \ref{fig:ebpf-throughput-8k}),
show the limitations of current network stacks. The hardware and operating
system can only drive 14-15 Gbps although the NIC supports 100 Gbps when running a single thread. 
The limitation arises because a single CPU can't drive packets fast enough through the kernel stacks,
which is consistent with prior works measuring UDP performance~\cite{cai2021understanding,zhang2021fastudp}.

In order to explore how \liha{} can scale to 100 Gbps speeds, we ran experiments using the DPDK drivers and
a user-space UDP stack on the FPGA platform, as shown in Figure~\ref{fig:DPDK-throughput}.
Recall the DPDK drivers allow a user-space
program to send and receive packets directly to the network interface card, bypassing the kernel.
DPDK is not applicable to the Userspace and eBPF platforms as they require operating system support. 
Figure~\ref{fig:DPDK-throughput} shows a \liha{} enabled NIC can drive 100 Gbps. This a positive result, as
often adding services to existing base IP protocols significantly degrades bandwidth performance
~\cite{liu2008passport}.

\subsection{Emulated DDoS attack}

\begin{figure}[t]
    \centering
    \includegraphics[width=0.5\textwidth]{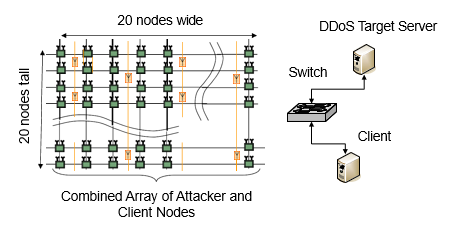}
    \caption{Emulated DDoS Attack Infrastructure.}
    \label{fig:DDoS_setup}
\end{figure}

We emulate a DDoS attack using a large scale emulation testbed. The testbed consists of a large array
of over 200 machines connected to a server as shown in Figure~\ref{fig:DDoS_setup}. 
Each machine in the array is connected to a switched network at 1 Gbps. A pair of machines a server
and client, are connected to the same switch at 100 Gbps.

The experiment uses the array machines as emulated attackers; each drives UDP packets at 1 Gbps at
the server. The client sends UDP to the server. The number of machines in the array, and thus the
bandwidth directed at the server, is the independent variable. The bandwidth of UDP traffic reaching the
server from the client is the dependent variable. As different \liha{} implementations have
different bandwidths, the metric we use is the fraction of the bandwidth under the no attack case
that reaches the server.

\begin{figure*}[t]
    \centering
    \begin{tabular*}{\textwidth}{@{\extracolsep{\fill}}cc}
      \includegraphics[width=0.5\textwidth]{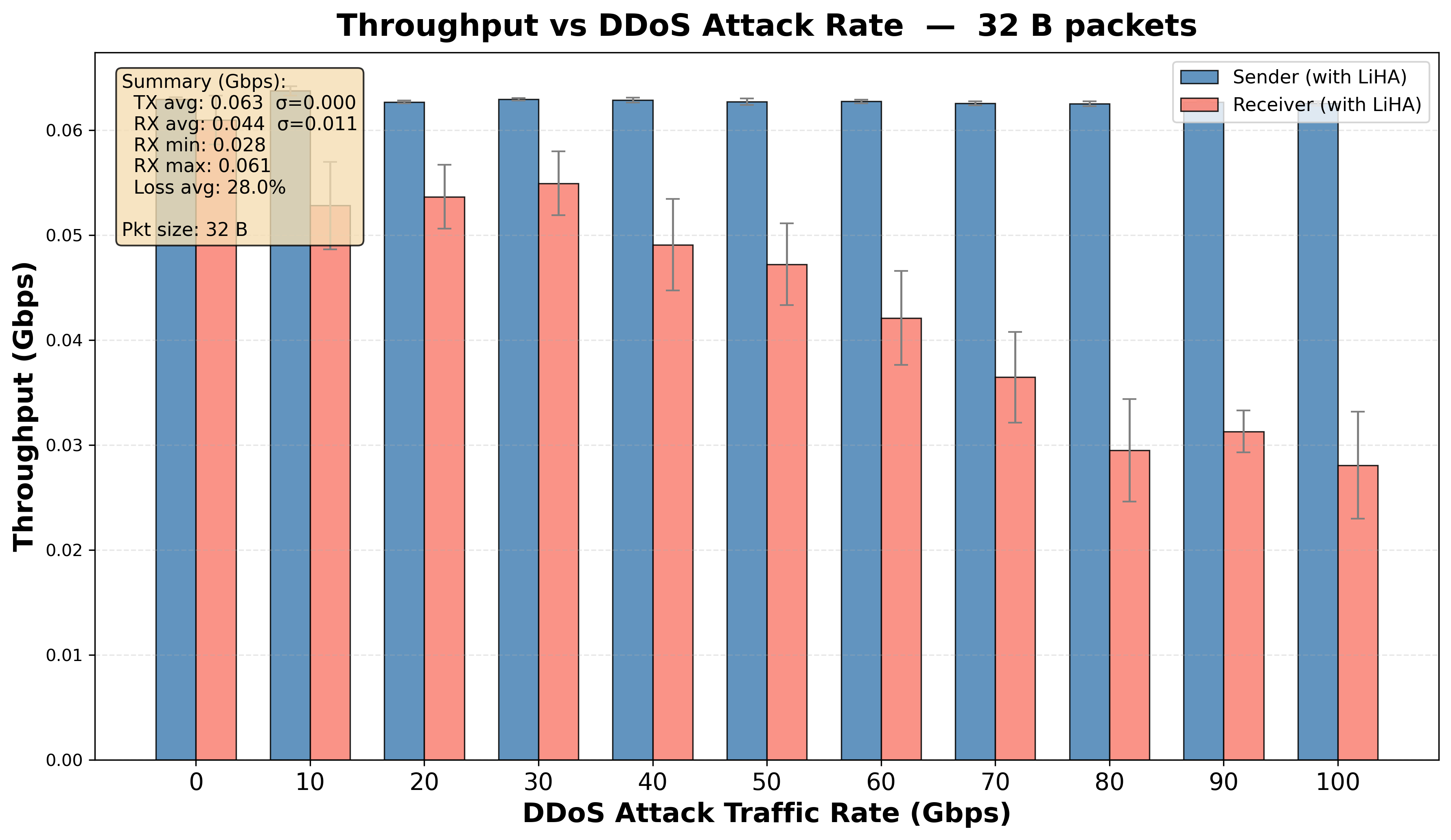} &
      \includegraphics[width=0.5\textwidth]{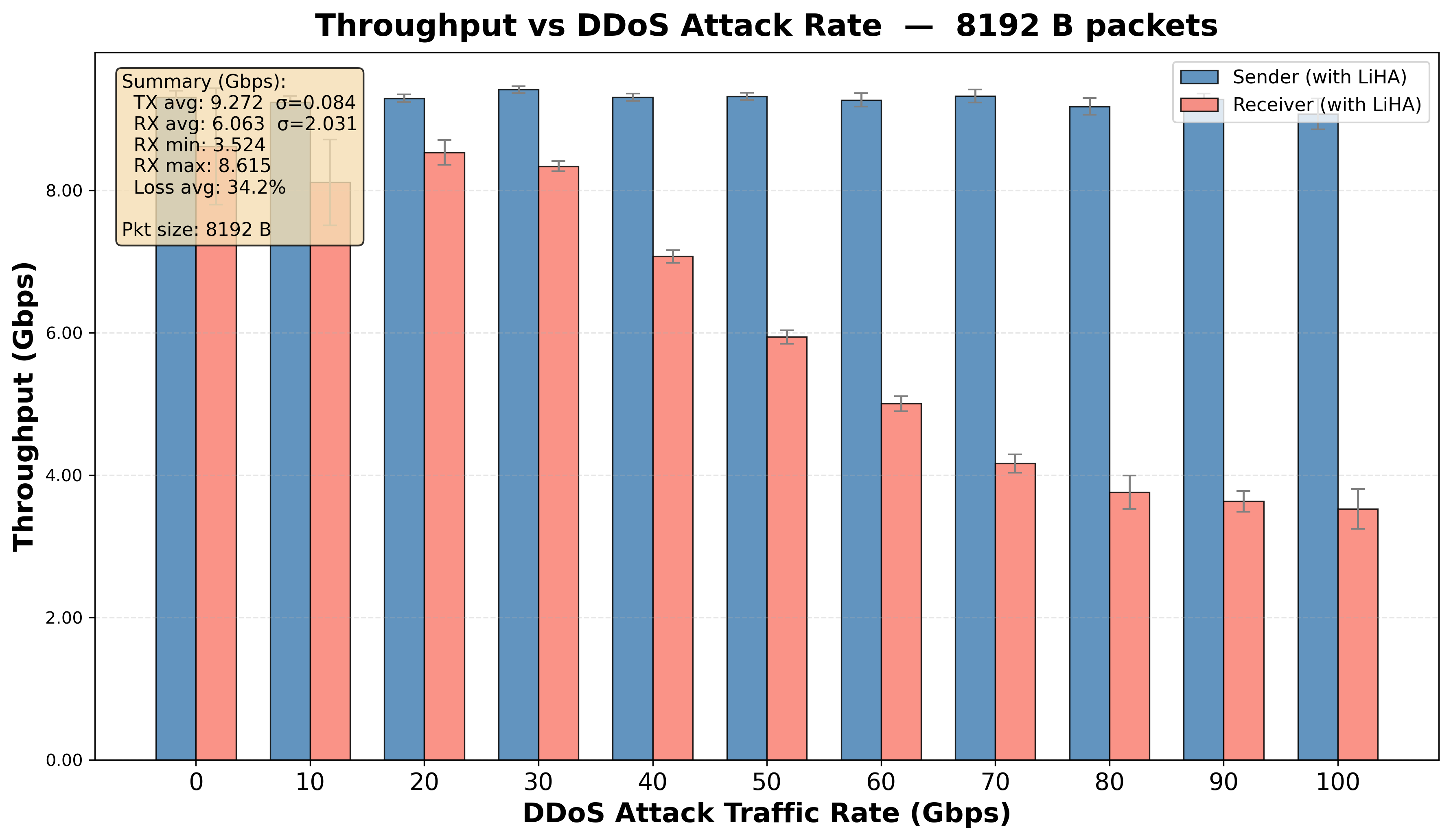} \\
      (A) & (B) \\
    \end{tabular*}
    \caption{DDoS bandwidth scaling with size 32 bytes of data packets (A) and 8192 byte jumbo frames (B).}
    \label{fig:ddos_all}
\end{figure*}

Figure~\ref{fig:ddos_all} shows the resulting bandwidth fractions
of the client's traffic getting through to the server for the three implementations. The figure shows
that for the Userspace implementation, the degradation profile is similar to that of eBPF, although as Figure~\ref{fig:all-throughput} shows, their baselines differ. Note that even when the DDoS attack scales to
saturation at 100Gbps, about 20\% of the legitimate stream can get through. We are investigating this effect, which
is likely due to difference in the queuing behavior of \liha{} packets from dropped DDoS ones. A \liha{} packet would have
a longer queue occupancy and interarrival spacing compared to DDoS traffic. Simple statistical models are unlikely
to describe these effects. 

The FPGA's bandwidth vs DDoS intensity profile is qualitatively different, with a plateau that reaches to about
80\% of the link bandwidth. Empirically dropping packets in the network interface results in more packets
getting through to the end host. While the additional client bandwidth by putting \liha{} in the NIC is not
surprising, the exact behavioral differences between packet types inside the NIC require more investigation.

The DDoS experiments show the effectiveness of different \liha{} implementations mitigating DDoS attacks,
although each one has a different level of performance. The DDoS experiment also shows the
portability of our approach;  \liha{}'s can be implemented at different cost/performance points in the network.
This allows network operators to perform incremental deployment based on the level of resources they have, they do
not need to change all their infrastructure at once.

\section{Development Challenges}
\label{sec:challenges}
We found the three implementations expose a clear
tradeoff between the required programmer expertise and the resulting
system performance. While userspace development is highly accessible
to traditional software engineers but yields the lowest throughput due
to overhead. eBPF demands specialized knowledge of kernel internals
and strict verifier rules to achieve near line-rate speeds. FPGA
design necessitates highly specialized digital hardware engineering
skills to realize line-rate performance.
\subsection{FPGA}
We found FPGA development presented significant complexity compared to the software implementations. Developers must manage
concurrency, clocked states, and strict timing constraints, while
grappling with long iteration cycles caused by full synthesis and
place and route recompilation. Tasks that differ in kind from software
development include architectural mapping, achieving timing closure
across heterogeneous pipelines, and ensuring cross-layer integration
between hardware DMA and kernel drivers. Functionally simulating an
algorithm does not guarantee successful hardware implementation;
designs must also meet synthesizability requirements and route
successfully without congestion. Furthermore, integrating inline
cryptographic hashing (like SHA-256) at sustained line-rate throughput
requires meticulous alignment with streaming packet boundaries,
careful buffer sizing for jumbo frames, and complex driver-level
coordination, making FPGA development for networking tasks a truly 
interdisciplinary systems engineering effort that requires deep
knowledge of both computer networks and HDL synthesis.
\subsection{Userspace}
This was the most accessible approach from the
programmer perspective. However, it still requires a deep knowledge of
low-level operating system interfaces, a complex networking stack,
and detailed knowledge of network protocols, which are waning skills in today's era. It utilizes
AF\_PACKET sockets and Linux traffic-control (TC) primitives to intercept and process packets,
circumventing the strict instruction limits of in-kernel
environments. While this approach allows the use of optimized
libraries like OpenSSL for SHA-256 computation, it introduces severe
challenges regarding packet interception, reinjection safety, and
bidirectional delivery. Reliably feeding traffic into user space
requires precise TC mirroring to an IFB device to avoid overloading
the application or capturing unintended frames. Furthermore,
developers must implement robust packet marking strategies to prevent
infinite reinjection loops where authenticated packets are repeatedly
intercepted. Handling bidirectional traffic adds another layer of
complexity, as inbound verified packets cannot be directly returned
via AF\_PACKET and must instead be carefully injected through a TUN
interface, demanding tight coordination between verification logic and
host-stack delivery.
\subsection{eBPF}
Much like the userspace implementation, the eBPF one
requires deep knowledge of the OS and networks. However, the overall
complexity to the userspace implementation was similar although the
programmer must be familiar with the eBPF environment, specifically
XDP for ingress and TC for egress. We found the primary development
challenge lies in satisfying the rigid safety requirements of the eBPF
verifier and staying within its 1-million instruction limit
constraint. Porting a complex algorithm like SHA-256 into eBPF's
"restricted C" requires extensive structural modifications, including
replacing dynamic memory allocation with BPF maps, implementing manual
memory operations, unrolling loops, and adding exhaustive bounds
checking. Because eBPF was designed for simple, high-performance
packet filtering rather than heavy cryptographic computation, adapting
such algorithms is arduous. Our experience is that adding complex
cryptography functions in the eBPF environment will require advanced
workarounds, such as offloading complex logic to custom kernel helper
functions (kfuncs), to successfully bypass eBPF's inherent limitations
for complex code.

\subsection{Development Summary}
Reflecting on the development process, each platform
presents a distinct profile of programmer effort and required
expertise. The userspace implementation is the most accessible for
traditional software engineers, leveraging standard libraries and
avoiding strict hardware or kernel verifier constraints, though it
demands a solid understanding of Linux traffic control and network
namespaces to avoid routing loops. eBPF development represents a steep
step up in difficulty, requiring specialized knowledge of kernel
internals, BPF maps, and restricted C programming to constantly battle
the rigid eBPF verifier. Finally, the FPGA implementation is by far
the most demanding and time-consuming, requiring highly specialized
digital design engineers who can reason about parallel clocked logic,
cross-domain hardware-software integration, and physical timing
closure, extending development iteration cycles from seconds to hours.

These varying development costs directly correlate with the
performance tradeoffs offered by each platform. Userspace development
is rapid and flexible but suffers from inherent performance
bottlenecks due to continuous kernel-to-userspace boundary crossings,
making it unsuitable for true high-speed forwarding. The eBPF approach
strikes a middle ground, offering excellent in-kernel performance that
approaches line rate for simpler operations, but its utility degrades
when burdened with computationally heavy tasks like SHA-256 hashing
without relying on external kernel functions. Conversely, the massive
investment in programmer time and specialized skills required for the
FPGA implementation yields unparalleled performance, successfully
delivering robust, deterministic line-rate authentication at 100 Gbps,
making it the only viable choice for carrier-grade core network
deployments.

\section{Conclusion}
\label{sec:conclusion}
The Internet still has no real source authentication mechanism. The problem is most proposed approaches are heavyweight and need extra work such as new headers, per-flow state, key exchange in data path; thus never deployed. \liha{} has a different view. It looks up keys based on source prefix using radix trie algorithm, then
computes one SHA-256 signature over selected IP header fields with the extracted key and embeds the results in the IP Options field of the header. It does not
perform any handshakes, contains no per-flow state, and doesn't introduce a new protocol. Key management is handled outside the forwarding path, so the data path has to do only two things: find a key and check the hash. 
The strength of \liha{} is its simplicity and minimalist design. This helps us to build \liha{} across three different environments. The user space was the easiest one to implement; yet the worst performance. We found that the eBPF implementation got reasonable single thread performance even though the encryption was in
software. An important result is that only the FPGA could get to 100 Gbps single thread performance, and that was only when totally bypassing the kernel.

\bibliographystyle{plain}
\bibliography{main}

\end{document}